\documentclass[prd,onecolumn,aps]{revtex4-2}
\usepackage{graphicx}
\usepackage{epsfig,graphics,subfigure,psfrag,amsmath,amssymb}
\usepackage{tabularx}
\usepackage{adjustbox}  
\usepackage{bm}
\usepackage{overpic}
\usepackage{xspace}
\usepackage{rotating}
\usepackage{color}
\usepackage{multirow}
\usepackage{colortbl}
\usepackage{epstopdf}
\usepackage{float}
\usepackage{lineno}
\usepackage{amsmath}
\usepackage{slashed,comment}

\usepackage{hyperref}  
\hypersetup{
    colorlinks=true,   %
    urlcolor=blue,     %
    citecolor=blue,     %
    linkcolor=blue,    %
}

\def\chicJ{\chi_{cJ}}
\def\chicj{\chicJ}
\def\bb {B\bar {B}}

\begin{document}

\graphicspath{{figure/}}
\DeclareGraphicsExtensions{.eps,.png,.ps}
\title{\boldmath Polarization analysis of $\chi_{cJ}$ decay into octet baryonic pairs }

\author{Cai-Ying Pang$^1$ }
\author{Rong-Gang Ping$^{2,4}$}
\email{pingrg@ihep.ac.cn}
\author{Dai-Hui Wei$^{3,1}$}
\email{weidh@gxnu.edu.cn}
\affiliation{$^1$School of Physical Science and Technology, Guangxi Normal University, Guilin 541004, China\\
$^2$Institute of High Energy Physics, Beijing 100049, People's Republic of China\\
$^3$Guilin Normal College, Guilin 541199, China\\
$^4$University of Chinese Academy of Sciences, Beijing 100049, China}


\begin{abstract}
This work presents a comprehensive analysis of polarization transfer in the decays \(\chi_{cJ}\) (\(J=0,1,2\)) to octet baryon-antibaryon pairs within a polarized electron-positron collision environment. Using the spin density matrix formalism, we trace the polarization from the initial beams through the production chain \(e^+e^- \to \psi(2S) \to \gamma \chi_{cJ}\) to the final-state baryon-antibaryon system. The helicity amplitude analysis for \(\chi_{c1} \to B\bar{B}\) confirms the universal angular distribution parameter \(\alpha = -1/3\), as dictated by the charge-conjugation helicity selection rule. For \(\chi_{c2}\) decays, \(\alpha\) and the transverse polarization depend on two independent amplitudes, and our quark-model calculations agree with existing data. We demonstrate that the longitudinal beam polarization \(P_z\) modifies the spin observables for \(\chi_{c1}\) and \(\chi_{c2}\), offering new experimental handles at future polarized facilities like the Super \(\tau\)-Charm Facility(STCF) to test decay mechanisms and explore baryonic spin entanglement as a quantum information resource.
\end{abstract}

\maketitle

\renewcommand{\arraystretch}{1.3}

\section{Introduction}

The study of charmonium physics provides a unique window into both perturbative and non-perturbative aspects of Quantum Chromodynamics (QCD). Among the charmonium states, the P-wave triplet $\chi_{cJ}$ ($J=0,1,2$) holds particular interest due to its intermediate mass scale, where the interplay between the production dynamics, governed by perturbative QCD (pQCD), and the subsequent hadronization into final-state hadrons, a non-perturbative process, can be intricately probed. The decay channels $\chi_{cJ} \to B\bar{B}$, where $B$ denotes a spin-1/2 baryon such as proton, neutron, or $\Lambda$, are especially valuable in this regard. These decays involve the creation of a baryon-antibaryon pair from a colorless, heavy quark-antiquark system, offering a testing ground for models of hadron formation, helicity conservation rules, and the role of higher-order QCD effects~\cite{Bolz:1997ez,Ma:2001ri,Liu:2010um}. Understanding the detailed decay mechanism of $\chi_{cJ}$ whether dominated by short-distance pQCD diagrams, influenced by non-relativistic QCD (NRQCD) matrix elements, or affected by rescattering and final-state interactions remains an active area of theoretical investigation. The polarization observables in these decays are particularly sensitive to the underlying dynamics, as they trace the transfer of angular momentum from the initial quarkonium state to the final-state baryons~\cite{BESIII:2018cnd,Perotti:2018wxm,Moortgat-Pick:2005jsx,Zhang:2025oks}.

Experimentally, significant progress has been made in measuring the branching fractions and angular distributions of $\chi_{cJ} \to B\bar{B}$ decays. Facilities such as the BESIII experiment at the Beijing Electron Positron Collider (BEPCII) have produced large and clean samples of $\chi_{cJ}$ particles via the radiative transition $\psi(2S) \to \gamma \chi_{cJ}$. Recent analyses have provided precise measurements of the branching fractions for channels like $\chi_{cJ} \to p\bar{p}$, $\Sigma^{0}\bar{\Sigma}^{0}$,$\Sigma^{+}\bar{\Sigma}^{-}$,$\Xi^-\bar{\Xi}^+$ and $\Xi^0\bar\Xi^0$  ~\cite{Bai2013chicj2BB,BESIII:chicjpp,Bai2003chicj2LL,BESIII:2025chicj2LL,Bai2020chicj2SS,BESIII:2022chicjXX}. Notably, these measurements reveal a distinct hierarchy in the decay rates among the different $\chi_{cJ}$ states and significant angular distribution parameters ($\alpha$ parameters) for the baryons, which are directly linked to the polarization transfer. The experimental data, with ever-increasing precision, pose both challenges and opportunities for theoretical models, calling for a more refined description that goes beyond the simplest pQCD or statistical hadronization pictures.

The polarization of the final-state baryons and the related phenomenon of spin entanglement constitute a frontier in quantum information aspects of particle physics~\cite{Horodecki:2009zz,Adesso:2007tx,Bernal:2024xhm}. In the decay $\chi_{cJ} \to B\bar{B}$, the spin-1/2 baryon and antibaryon pair is produced in a quantum-correlated state. For the $\chi_{c1}$ and $\chi_{c2}$ decays, the initial state possesses nonzero spin alignment, which can lead to a non-trivial spin density matrix for the baryon-antibaryon system~. This entanglement, a direct consequence of quantum mechanics and angular momentum conservation, can be quantified through observables like spin correlations and entanglement entropy\cite{Hong:2025drg,Li:2026bkf}. Recent theoretical works have begun to explore these quantum correlations in heavy quarkonium decays, suggesting that such processes could serve as high-energy laboratories for testing fundamental quantum principles\cite{ATLAS:2023fsd,Barr:2024djo,Dong:2023xiw}. However, most existing analyses assume an unpolarized initial $\chi_{cJ}$ state, typically produced from unpolarized $e^+e^-$ collisions. The role of the initial production mechanism in shaping the final spin correlations remains less explored.

In this work, we extend the polarization analysis  of $\chi_{cJ} \to B\bar{B}$ decays, with $B$ representing octet baryon, to a more general and experimentally relevant scenario: the production of $\chi_{cJ}$ from the annihilation of polarized electron and positron beams. Polarized beam experiments, such as those possible at future $e^+e^-$ colliders like the STCF~\cite{Achasov:2023gey} or upgraded BEPCII, offer an additional handle to control and manipulate the initial angular momentum state. By considering polarized initial beams, we introduce a new axis of quantization and a controllable source of spin alignment for the $\chi_{cJ}$ meson. This allows us to study how the beam polarization propagates through the production process ($e^+e^- \to \psi(2S)\to\gamma\chi_{cJ}$) and subsequently transfers to the final-state baryons via the decay $\chi_{cJ} \to B\bar{B}$. We systematically calculate the complete set of helicity amplitudes for these processes within a quark-model framework, incorporating both the production and decay dynamics. Our analysis aims to predict the resulting baryon spin polarizations and spin correlations as functions of the beam polarization parameters. We quantify the expected experimental observables, such as angular distributions and polarization correlation, and discuss how measurements of these observables with polarized beams could provide unprecedented constraints on the $\chi_{cJ}$ decay mechanism, test helicity selection rules with greater discriminative power, and offer new insights into the spin entanglement properties of the baryon-antibaryon system. This study bridges the gap between traditional charmonium spectroscopy and the emerging field of high-energy quantum information, highlighting the potential of polarized colliders as powerful tools for fundamental QCD studies.

\section{spin density matrix}

\subsection{$\psi(2S)$ particle}

\begin{figure}[htbp]
\centering
\includegraphics[width=0.4\textwidth]{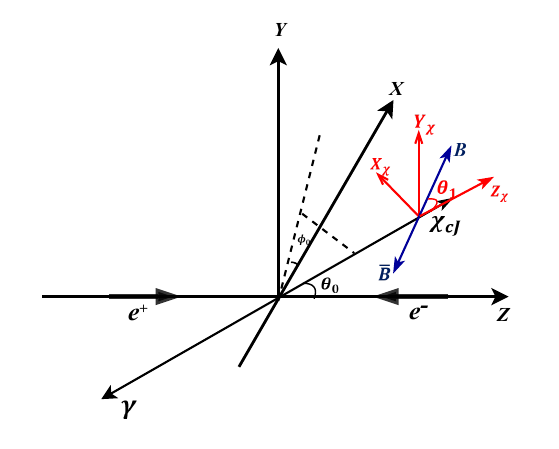}
\caption{ Definition of helicity system and helicity angles  for
$\psi(2S) \to \gamma \chi_{cJ}$,$\chi_{cJ} \to B\bar{B}$.}
\label{frame}
\end{figure}

In $ e^+ e^- $ colliders, the laboratory coordinate system is conventionally defined with the $ z $-axis aligned along the beam direction, typically chosen as the momentum direction of the positron ($ e^+ $). The $ x $-axis is usually set perpendicular to the beam axis within the accelerator plane, and the $ y $-axis completes the right-handed Cartesian system (as shown in Fig.~\ref{frame}). When the electron beams possess both longitudinal and transverse polarizations, the spin density matrix (SDM) for the initial $ e^+ e^- $ system must account for these polarization states. For an electron with longitudinal polarization $ P_z $ and transverse polarization $ \vec{P}_t = (P_x, P_y)=P_T e^{i\phi_e} $, with $\phi_e$  the azimuthal angle, its individual spin density matrix can be expressed as $ \rho_e = \frac{1}{2}(I + \vec{P} \cdot \vec{\sigma}) $, where $ \vec{P} = (P_x, P_y, P_z) $ and $ \vec{\sigma} $ are the Pauli matrices. The combined SDM for the colliding $ e^+ e^- $ pair is then the direct product of their individual matrices: $ \rho_{e^+ e^-} = \rho_{e^+} \otimes \rho_{e^-} $.

In the process $ e^+ e^- \rightarrow \psi(2S) $, assuming helicity conservation and parity conservation in the coupling via a virtual photon, the helicity of the virtual photon is restricted to $ \lambda_\gamma = \pm 1 $. The SDM for the produced $ \psi(2S) $ can then be derived from the decay matrix formalism. The general relation is $ \rho_f = M \rho_i M^\dagger $, where $ \rho_i $ is the SDM of the initial $ e^+ e^- $ system, and $ M $ is the decay amplitude matrix for $ e^+ e^- \rightarrow \gamma^* \rightarrow \psi(2S) $, expressed in terms of helicity amplitudes and Wigner $ D $-functions. 
For the specific case discussed in this paper, which focuses on the longitudinal polarization $P_z(\bar{P}_z)$ of the $e^-$ and $e^+$ beams, the transverse polarization $P_T$ of both beams also arises from the self-polarization due to the Sokolov-Ternov effect~\cite{Jackson:1975qi}.
Thus one has the SDM for $\psi(2S)$ particle :

\begin{equation}\label{Eq:psisdm}
\rho^{\gamma*/\psi(2S)} = \frac{1}{2}
\begin{pmatrix}
(1+\bar P_z)(1-P_z) & 0 & P_T^2 \\
0 & 0 & 0 \\
P_T^2 & 0 & (1-\bar P_z)(1+P_z)
\end{pmatrix}.
\end{equation}
At the $\psi(2S)$ production energy, the transverse polarization $P_T$ in BEPCII reaches $0.28$ after one hour of beam injection, following the Sokolov-Ternov mechanism with a characteristic rise time $\tau_0=2.8$ hours~\cite{Cao:2024tvz}. The polarization magnitude increases with beam energy under identical injection conditions, demonstrating the energy-dependent polarization buildup.

\subsection{$\chicJ$ states}
In the decay $\psi(2S)(\lambda) \to \chi_{cJ}(\lambda_1)\gamma(\lambda_2)$, the helicity angle $\theta_0$ is defined as the angle between the $\chi_{cJ}$ momentum and the positron direction in the $\psi(2S)$ rest frame, and the azimuthal is denoted by $\phi_0$. The helicity amplitudes $A^{(J)}_{\lambda_1,\lambda_2}$ describe the decay amplitude for specific helicity states $\lambda_1$ (of $\chi_{cJ}$) and $\lambda_2$ (of $\gamma$), where $\lambda_i$ denotes the spin projection along the particle's momentum direction.

This charmonium transition conserves parity, leading to specific symmetry relations among the helicity amplitudes. 
For the $\chi_{c0}$ final state, the amplitudes satisfy $A^{(0)}_{0,-1} = A^{(0)}_{0,1}$. In the case of $\chi_{c1}$, the relations are $A^{(1)}_{-1,-1} = -A^{(1)}_{1,1}$ and $A^{(1)}_{0,-1} = -A^{(1)}_{0,1}$. For $\chi_{c2}$, the amplitudes follow $A^{(2)}_{-2,-1} = A^{(2)}_{2,1}$, $A^{(2)}_{-1,-1} = A^{(2)}_{1,1}$, and $A^{(2)}_{0,-1} = A^{(2)}_{0,1}$. These constraints arise from parity conservation and the spin properties of the final states~\cite{Tabakin:1985yv}.

The helicity amplitude $A^{(J)}_{\lambda_1,\lambda_2}$ encodes the dynamical information for the transition $\psi(2S) \to \chi_{cJ}\gamma$. Theoretical arguments and BESIII measurements confirm that this process is dominated by electric dipole (E1) transitions~\cite{BESIII:2011nst}. Consequently, the helicity amplitudes follow E1 selection rules\cite{Karl:1975jp}: for $\chi_{c1}$, $A^{(1)}_{1,1} = A^{(1)}_{0,1}$; for $\chi_{c2}$, $A^{(2)}_{2,1} = \sqrt{2}A^{(2)}_{1,1} = \sqrt{6}A^{(2)}_{0,1}$. These relations reflect the angular momentum and parity constraints of E1 transitions.

The spin density matrix (SDM) of $\chi_{cJ}$ can be expressed as a linear transformation of the $\psi(2S)$ SDM:
\begin{equation}\label{Eq:chcjsdm}
\rho(\chi_{cJ}) = M \rho^{\gamma*/\psi(2S)} M^\dagger, \quad \text{with} \quad M_{\lambda,\lambda_1-\lambda_2} = \mathcal{N}_J D^{1}_{\lambda,\lambda_1-\lambda_2}(\phi_0,\theta_0,0) A^{(J)}_{\lambda_1,\lambda_2}.
\end{equation}
Here, $\mathcal{N}_J$ is a normalization factor, and $D^{1}_{i,j}$ denotes the Wigner $D$-matrix element. To note that the azimuthal angle $\phi_0$ will yield the dependence of the polarization of $P_T$ and $P_z$. 
The elements of the spin density matrix $\rho(\chi_{cJ})$ can be written as
$
\rho_{\lambda_1, \lambda'_1}(\chi_{cJ}) \propto  \sum_{\lambda, \lambda',  \lambda_2} \rho_{\lambda, \lambda'}^{\psi(2S)}  D_{\lambda, \lambda_1 - \lambda_2}^{1} (\Omega_0)  D_{\lambda', \lambda'_1 - \lambda_2}^{1*} (\Omega_0)
\times A_{\lambda_1, \lambda_2}^{(J)} A_{\lambda'_1, \lambda_2}^{(J)*}, 
$
where $\lambda(\lambda')=\pm1$ denote the helicities of the $\psi(2S)$ produced in $e^+e^-$ annihilation,  and $\lambda_2=\pm1$ corresponds to the photon helicity.
This expression explicitly shows how the polarization components of the $\psi(2S)$, encoded in $\rho_{\lambda\lambda'}^{\psi(2S)}$, are mapped onto the $\chi_{cJ}$ states through the radiative transition.

Combining Eq.~(\ref{Eq:psisdm}) and Eq.~(\ref{Eq:chcjsdm}), using the E1 selection rules, the angular distribution of $\chicJ$ particle is obtained by take trace of its SDM, 
\begin{equation}
{d\sigma \over d\cos\theta_0 d\phi_0}\propto
\left\{
\begin{array}{lc}
1 & \text{for $\chi_{c0}$,}\\
\frac{1}{2} \left[ (1 - \bar{P}_zP_z) (5 - \cos2\theta_0) - 2 P_T^2 \cos2\phi_0 \sin^2\theta_0 \right] |A_{0,1}|^2 & \text{for $\chi_{c1}$},\\
 \frac{1}{12} \left[ (1 - \bar{P}_z P_z) (27 + \cos2\theta_0) + 2 P_T^2 \cos2\phi_0 \sin^2\theta_0 \right] |A_{2,1}|^2& \text{for $\chi_{c2}$}.
\end{array}
\right.
\end{equation}
One can see that the $\phi_0$ distribution depends on $P_T$, which provides a means to measure the beam transverse polarization. Moreover, if the positron is unpolarized, the $\cos\theta_0$ distribution is independent of the electron polarization. For unpolarized beams $(\bar{P}_z = P_z = P_T = 0)$, the angular distribution reduces to ${d\sigma}/{d\cos\theta_0}\propto 1+\alpha\cos^2\theta_0$, with $\alpha=0, -1/3, 1/13$ for $\chi_{c0}$, $\chi_{c1}$, and $\chi_{c2}$~\cite{ChenPing2020}, respectively.

From Eq.~\eqref{Eq:chcjsdm}, it is evident that the spin density matrix of $\chi_{cJ}$ depends on the helicity angles $\theta_0$ and $\phi_0$. To investigate the polarization effects in the subsequent decay $\chi_{cJ} \to B\bar{B}$ through baryon angular distributions, we adopt the momentum-space averaged SDM of $\chi_{cJ}$. This is obtained by integrating over $\theta_0$ and $\phi_0$, effectively marginalizing the angular dependence. If E1 approximation is assumed, one has the normalized SDM of $\chi_{c1}$
\begin{equation}
\mathcal{N}_1\rho(\chi_{c1})=
\begin{pmatrix}
-16 (-1 + \bar P_z P_z) & -3 \sqrt{2} \pi (\bar P_z - P_z) & 0 \\
-3 \sqrt{2} \pi (\bar P_z - P_z) & -32 (-1 + \bar P_z P_z) & -3 \sqrt{2} \pi (\bar P_z - P_z) \\
0 & -3 \sqrt{2} \pi (\bar P_z - P_z) & -16 (-1 + \bar P_z P_z)
\end{pmatrix}
,
\end{equation}
where the normalization factor $\mathcal{N}_1=64 ( 1-\bar P_{z} P_z) $. For $\chi_{c2}$ state, one has the normalized SDM of $\chi_{c2}$

\begin{equation}
\mathcal{N}_2 \rho(\chi_{c2})=
\begin{pmatrix}
-96 (1 - \bar{P}_z P_z) & 18 \pi (-\bar{P}_z + P_z) & -8 \sqrt{6} (1 - \bar{P}_z P_z) & 0 & 0 \\
18 \pi (-\bar{P}_z + P_z) & -48 (1 - \bar{P}_z P_z) & -3 \sqrt{6} \pi (\bar{P}_z - P_z) & 0 & 0 \\
-8 \sqrt{6} (1 - \bar{P}_z P_z) & -3 \sqrt{6} \pi (\bar{P}_z - P_z) & -32 (1 - \bar{P}_z P_z) & -3 \sqrt{6} \pi (\bar{P}_z - P_z) & -8 \sqrt{6} (1 - \bar{P}_z P_z) \\
0 & 0 & -3 \sqrt{6} \pi (\bar{P}_z - P_z) & -48 (1 - \bar{P}_z P_z) & 18 \pi (-\bar{P}_z + P_z) \\
0 & 0 & -8 \sqrt{6} (1 - \bar{P}_z P_z) & 18 \pi (-\bar{P}_z + P_z) & -96 (1 - \bar{P}_z P_z)
\end{pmatrix}
,
\end{equation}
the normalization factor $\mathcal{N}_2=320 ( -1+ \bar{P}z P_z) $.
The averaged SDM then serves as the input for analyzing the $B\bar{B}$ decay angular correlations, where the initial $\chi_{cJ}$ polarization is imprinted on the final-state baryon pair. This approach isolates the intrinsic polarization transfer from the $\chi_{cJ}$ production dynamics, independent of the $\psi(2S) \to \chi_{cJ}\gamma$ kinematic configuration.

\subsection{$B\bar B$ pair}
In the decay process \(\chi_{cJ}(\lambda_1)\to B(\lambda_3)\bar{B}(\lambda_4)\), we investigate polarization effects in the helicity frame of the \(B\bar{B}\) system. The helicity frame is constructed starting from the center-of-mass frame of the \(\chi_{cJ}\). In this frame, the baryon and antibaryon are emitted back-to-back. The momentum direction of the baryon is chosen as the \(Z_B\)-axis of the \(B\bar{B}\) helicity frame. The \(Y_B\)-axis is taken to be perpendicular to the plane spanned by the momentum of the \(\chi_{cJ}\) and that of the baryon, while the \(X_B\)-axis is defined such that \((X_B, Y_B, Z_B)\) forms a right-handed coordinate system, as illustrated Fig.~\ref{frame}. In this frame, the helicity amplitude of the \(B\bar{B}\) pair is denoted by \(B^{J}_{\lambda_3,\lambda_4}\), and the spin density matrix of the two-body system can be expressed as:
\begin{eqnarray}
\rho^{(J)}(BB)&=& M\rho(\chicj) M^\dagger,\nonumber \text{~~ with~} \\ M_{\lambda_1,\lambda_3-\lambda_4}&=&D^J_{\lambda_1,\lambda_3-\lambda_4}(\phi_1,\theta_1,0) B^{(J)}_{\lambda_3,\lambda_4},
\end{eqnarray}
where $J=0,1$ and $2$ for $\chi_{c0},\chi_{c1}$ and $\chi_{c2}$ states.

For the decay \(\chi_{cJ} \to B\bar{B}\), parity conservation implies the following relations for the helicity amplitudes~\cite{Tabakin:1985yv}:
\(
B^{(J)}_{-\lambda_3,-\lambda_4} = (-1)^J \, B^{(J)}_{\lambda_3,\lambda_4},
\)
which gives
\(
B^{(0,2)}_{-\lambda_3,-\lambda_4}=B^{(0,2)}_{\lambda_3,\lambda_4}  \text{~for } \chi_{c0,2},
B^{(1)}_{-\lambda_3,-\lambda_4}=-B^{(1)}_{\lambda_3,\lambda_4}  \text{~for } \chi_{c1}.
\)
In addition, the charge conjugation symmetry requires
\(
B^{(1)}_{\lambda_3,\lambda_4} = -B^{(1)}_{\lambda_4,\lambda_3},
\)
so that
\(
B^{(1)}_{-1/2,-1/2}=B^{(1)}_{1/2,1/2}=0.
\)
Consequently, only the amplitudes \(B^{(1)}_{-1/2,1/2}\) and \(B^{(1)}_{1/2,-1/2}\) contribute to the decay \(\chi_{c1}\to B\bar{B}\), with the relation
\(
B^{(1)}_{-1/2,1/2}= -\,B^{(1)}_{1/2,-1/2}.
\)
This helicity selection rule implies that the helicity amplitude for $\chi_{c1} \to B\bar{B}$ is governed by a single parameter. For the $\chi_{c2}$ decay, the two independent amplitudes are  parameterized by their phase difference $\Delta$:
\(
B^{(2)}_{-\frac{1}{2},\frac{1}{2}} =  e^{i\Delta} B^{(2)}_{\frac{1}{2},\frac{1}{2}}.
\)

\subsubsection{$\chi_{c0}\to\bb$}
With these relation, one has the SDM for $\chi_{c0}$ decays in terms of helicity basis
\begin{equation}
\label{rho0bb}
\rho^{(0)}(\bb)={1\over 2}
\begin{pmatrix}
1 & 0 & 0 & 1 \\
0 & 0 & 0 & 0 \\
0 & 0 & 0 & 0 \\
1 & 0 & 0 & 1
\end{pmatrix}.
\end{equation}
It follows that the baryons angular distribution is isotropic in the $\bb$ helicity frame. 
\subsubsection{$\chi_{c1}\to\bb$}
While for $\chi_{c1}$ decay, the SDM of the $\bb$ reads
\begin{equation}
\rho^{(1)}(\bb)\propto
\begin{pmatrix}
0 & 0 & 0 & 0 \\
0 & \rho_{22} & \rho_{23} & 0 \\
0 & \rho_{32} & \rho_{33} & 0 \\
0 & 0 & 0 & 0
\end{pmatrix},
\end{equation}
with
\begin{eqnarray}
\label{rho1BB}
\rho_{22}&=&2 \left[ 2 (\bar P_z P_z - 1) (\cos 2\theta_1 - 5)
+ 3\pi (P_z - \bar P_z) \cos\phi_1 \sin\theta_1 \right]
| B_{\frac{1}{2}, -\frac{1}{2}} |^2,\nonumber\\
\rho_{33}&=&2 \left[ 2 (\bar P_z P_z - 1) (\cos 2\theta_1 - 5)
+ 3\pi (\bar P_z -  P_z) \cos\phi_1 \sin\theta_1 \right]
| B_{\frac{1}{2}, -\frac{1}{2}} |^2,\\
\rho_{23}&=&\rho_{32}=8 (1- \bar P{z} P_z  ) \sin^2\theta_1
| B_{\frac{1}{2},-\frac{1}{2}} |^2.\nonumber
\end{eqnarray}
It follows that the baryons angular distribution in the $\bb$ helicity frame is expressed
\begin{equation}\label{chic1:alpha}    
I^{(1)}(\theta_1)\propto  (1-\bar P_z P_z ) (5-\cos 2\theta_1 )| B_{\frac{1}{2},-\frac{1}{2}} |^2.
\end{equation}
The angular distribution is given by $1-\frac{1}{3}\cos^2\theta_1$, a form independent of the beam polarization.
\subsubsection{$\chi_{c2}\to\bb$}
After the integration of $\phi_1$, the spin density of $\chi_{c2} $ reads
\begin{equation}
\label{rho2bb}
\rho^{(2)}(\bb)\propto
\begin{pmatrix}
2e^{i\Delta}(3c_{2\theta}-7)b_{++}^2 & \sqrt{6}s_{2\theta}b_{+-}b_{++} & -\sqrt{6}s_{2\theta}b_{+-}b_{++} & 2e^{i\Delta}(3c_{2\theta}-7)b_{++}^2 \\
\sqrt{6}e^{2i\Delta}s_{2\theta}b_{+-}b_{++} & 3e^{i\Delta}(c_{2\theta}-5)b_{+-}^2 & -6e^{i\Delta}s_{\theta}^2b_{+-}^2 & \sqrt{6}e^{2i\Delta}s_{2\theta}b_{+-}b_{++} \\
-\sqrt{6}e^{2i\Delta}s_{2\theta}b_{+-}b_{++} & -6e^{i\Delta}s_{\theta}^2b_{+-}^2 & 3e^{i\Delta}(c_{2\theta}-5)b_{+-}^2 & -\sqrt{6}e^{2i\Delta}s_{2\theta}b_{+-}b_{++} \\
2e^{i\Delta}(3c_{2\theta}-7)b_{++}^2 & \sqrt{6}s_{2\theta}b_{+-}b_{++} & -\sqrt{6}s_{2\theta}b_{+-}b_{++} & 2e^{i\Delta}(3c_{2\theta}-7)b_{++}^2
\end{pmatrix}\cdot ( \bar P_z Pz -1 )e^{-i\Delta},
\end{equation}
where $c_{2\theta} = \cos 2\theta_1$, $s_{2\theta} = \sin 2\theta_1$, $s_{\theta} = \sin\theta_1$, $b_{++} = |B_{1/2,1/2}|$, $b_{+-} = |B_{1/2,-1/2}|$, and $\Delta$ is the phase angle difference between $B_{1/2,-1/2}$ and $B_{1/2,1/2}$ .

It follows that the baryon's angular distribution in the $\bb$ helicity frame is expressed
\begin{equation}
I^{(2)}(\theta_1)\propto  (1-\bar P_z P_z ) (1+\alpha\cos^2\theta_1 )\text {~with ~} \alpha = - {3b^2_{+-}+6b_{++}^2 \over 9b^2_{+-}+10b_{++}^2}.
\label{chic2 angur}
\end{equation}
To note that in the approximation of E1 transition, the angular distribution is independent on the $\cos^4\theta_1$ term.

\section{ Polarization $\bb$ in $\chi_{cJ}$ decays }
Spin entanglement between baryon pairs is a paradigmatic manifestation of non-classical correlations in spin systems and has become an important resource in quantum physics. It not only provides a crucial experimental platform for testing the foundations of quantum mechanics -- such as Bell nonlocality and local realism, but is also being extended as a sensitive quantum probe for detecting new physics. Within this framework, spin polarization serves as an ideal physical carrier for realizing and manipulating baryonic entanglement, making its preparation and measurement techniques central to experimental research. For a two-body spin-1/2 system, its complete quantum information can be fully described by the respective baryon(antibaryon) polarization vectors $O_i (\bar O_j)(i,j=x,y,z)$ and the second-order spin correlation tensor $C_{ij} = \langle \sigma_i \otimes \sigma_j \rangle$. In terms of Pauli matrices, the spin density matrix for $\bb$ reads
\begin{equation}\label{rhoBB}
\rho^{(J)}(\bb)={1\over 4}[\mathbf{I}\otimes \mathbf{I}+ \sum_i{O_i}(\sigma_i\otimes \mathbf{I}) + \sum_j \bar O_j(\mathbf{I}\otimes \sigma_j)+\sum_{ij}C_{ij}\sigma_i\otimes \sigma_j],
\end{equation}
where denote the $2\times 2$ identity matrix.

On the other hand, the  polarization vector and the correlation tensor can be expressed if the SDM is available with
\begin{eqnarray}\label{polar_corr}
O_i &=& Tr[\rho^{(J)}(\bb) \sigma_i\otimes \mathbf{I}],\nonumber\\
\bar O_j &=& Tr[\rho^{(J)}(\bb) \mathbf{I}\otimes \sigma_i],\\
C_{ij} &=& Tr[\rho^{(J)}(\bb)\sigma_i\otimes  \sigma_j ]\nonumber.
\end{eqnarray}

\subsubsection{$\chi_{c0}\to \bb$}
Using the SDM for $\chi_{c0}\to \bb$ as given by Eq.~(\ref{rho0bb}), we get the polarization vector $O_i=\bar O_i=0~(i=x,y,z)$ for baryon and antibaryon. The spin correlation matrix $C_{ij}$ defined in Eqs.~(\ref{rhoBB},\ref{polar_corr}) reads
\begin{equation}
C=\begin{pmatrix}
1&0&0\\
0&-1&0\\
0&0&1
\end{pmatrix}.
\end{equation}

The \(\chi_{c0} \to B\bar{B}\) decay deserves special attention for quantum information studies. For a bipartite \(B\bar{B}\) system described by the density matrix \(\rho^J(B\bar{B})\), the entanglement of formation is measured by the concurrence \(\mathcal{C}[\rho]\)~\cite{Wootters:1997id}. Following the standard definition, \(\mathcal{C}[\rho] = \max(0, t_1 - t_2 - t_3 - t_4)\), with \(t_i\) being the square roots of the eigenvalues of \(R = \rho(\sigma_y \otimes \sigma_y)\rho^*(\sigma_y \otimes \sigma_y)\) arranged in decreasing order. With the joint spin density matrix \(\rho^{B\bar{B}}\), we find \(\mathcal{C}[\rho] = 1\) for the \(\chi_{c0}\) decay, which implies that the \(\chi_{c0}\) state is maximally entangled~\cite{Hong:2025drg}.

\subsubsection{$\chi_{c1}\to \bb$}
Using the SDM for $\chi_{c1}\to \bb$, one gets the polarization vector $O_i=\bar O_i=0~(i=x,y)$ and $O_z=-\bar O_z={\rho_{33}-\rho_{22}\over \rho_{22}+\rho_{33}}$ for baryon and antibaryon. To note that the longitudinal polarization is useful to measure the hyperon further decay. The spin correlation matrix 
$C_{ij}$ defined in Eqs.~(\ref{rhoBB},\ref{polar_corr}) reads
\begin{equation}
\label{chic1_Cij}
C={1\over \rho_{22}+\rho_{33}}\begin{pmatrix}
\rho_{23}+\rho_{32}&i(\rho_{32}-\rho_{23})&0\\
i(\rho_{23}-\rho_{32})&\rho_{23}+\rho_{32}&0\\
0&0&-1
\end{pmatrix},
\end{equation}
here $\rho_{ij}(i,j=1,2,3)$ are the SDM elements of the the $\bb$ for $\chi_{c1}$ decay, as given by Eq.~(\ref{rho1BB}).

\subsubsection{$\chi_{c2}\to \bb$}
For $\chi_{c2}\to \bb$, using the Eq.~(\ref{rho2bb}), the polarization vectors for the baryon and antibaryon are obtained as $O_i=\bar O_i=0~(i=x,z)$, and
\begin{equation}\label{chic2_O_y}
O_y=-\bar O_y={1\over \mathcal{F}}(4\sqrt{6} \cos\theta_1 \sin\theta_1 \sin\Delta \, b_{+-} b_{++}),
\end{equation} 
The spin correlation matrix reads
\begin{eqnarray}
C_{11}&=&{1\over \mathcal{F}}[2(3\cos 2\theta_1 - 7) b_{++}^2 - 3(1 - \cos 2\theta_1) b_{+-}^2],\\
C_{22}&=&{1\over \mathcal{F}}[2 \bigl(7 - 3\cos 2\theta_1\bigr) b_{++}^2 - 6 \sin^2\theta_1 \, b_{+-}^2],\\
C_{33}&=&{1\over \mathcal{F}}[-3 (\cos 2\theta_1 - 5) b_{+-}^2 + 2 (3\cos 2\theta_1 - 7) b_{++}^2],\\
C_{31}&=&-C_{13}={1\over \mathcal{F}}[4\sqrt{6} \cos\Delta \cos\theta_1 \sin\theta_1 \, b_{+-} b_{++}],\\
C_{12}&=&C_{21}=C_{23}=C_{32}=0,\label{chic2_C_ij}
\end{eqnarray}
with
\[
\mathcal{F}=3 (\cos 2\theta_1 - 5) b_{+-}^2 + 2 (3\cos 2\theta_1 - 7) b_{++}^2.
\]

\section{ Helicity amplitude in quark model }
\begin{figure}[htbp]
\centering
\includegraphics[width=0.3\textwidth]{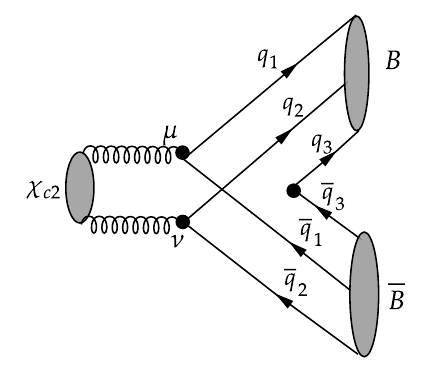}
\caption{ $\chi_{c2}$ decays into baryon antibaryon pair.}
\label{fig:tree}
\end{figure}

We compute the helicity amplitudes for $\chi_{c2} \to B\bar B$ using the conventional constituent quark model~\cite{Ping:2004sh,Ping:2002,3P:1996yt}. At tree level (see Fig.~\ref{fig:tree}), the $c\bar c$ pair in the $\chi_{c2}$ annihilates into two gluons, which subsequently convert into two $q\bar q$ pairs and combine with an additional $q\bar q$ pair (with $0^{++}$ quantum numbers) produced from the QCD vacuum. The three (anti)quarks then hadronize into (anti)baryons, forming the final-state baryon-antibaryon pair. As an approximation, only the contribution of the baryon wave function at the origin in momentum space is taken into account; that is, the momentum of each constituent quark is taken as a fixed fraction of the baryon momentum, weighted by the mass of the quark that constitutes the baryon.

In estimating the angular distribution coefficients for $\chi_{c2} \to B\bar B$ using helicity amplitudes, the decay mechanism described by the constituent quark model at tree level can be further simplified. Considering Eq.~(\ref{chic2 angur}), the angular distribution coefficients are determined by the ratios of helicity amplitudes. The QCD nonperturbative effects introduced by the initial- and final-state hadron wave functions, as well as common color factors, cancel out in these ratios, leaving only the dependence on the helicities of the quark pairs. These ratios of helicity amplitudes constitute the primary factor in determining the angular distribution coefficients. The helicity amplitudes for the decay $\chi_{c2} \to B\bar B$ can be written as

\begin{align}
b_{\lambda_3,\lambda_4} &=F\langle \Psi_B\Psi_{\bar B} |  \bar{u}(\lambda_1) \gamma^{\mu} v(\bar{\lambda}_1) \, \bar{u}(\lambda_2) \gamma^{\nu} v(\bar{\lambda}_2) \bar{u}(\lambda_3) v(\bar{\lambda}_3)  \nonumber\\
&+(u\leftrightarrow s)+(d\leftrightarrow s) |\epsilon_{\mu\nu}(\lambda_3-\lambda_4)\rangle,
\label{chic2_amp}
\end{align}
where $F$ is an overall constant that accounts for the initial- and final- state hadron wave functions at the origin, as well as the color factor. Here $u(v)$ denotes the free Dirac spinor for a quark (antiquark), $\Psi_B$ ($\Psi_{\bar B}$) is the spin wave function of the baryon (antibaryon), including the spin $\chi_B(\chi_{\bar B})$ and flavor wave functions ($\phi_B$),  and $\epsilon_{\mu\nu}(\lambda_3-\lambda_4)$ represents the spin wave function of the $\chi_{c2}$ with helicity $\lambda_3-\lambda_4$. Their explicit expressions of wave functions are given in appendix. In  the  constituent quark model,  we parameterize nonperturbative QCD effects, including hadronic bound states and long-distance effects, into a single overall constant.
Their values can be determined from the measured decay width. However, they cancel out in the calculation of helicity amplitude ratios.

We calculate the amplitude in the $\bb$ helicity system, where the spin quantization axis ($Z$-axis) is chosen along the baryon momentum direction, placing the antibaryon along the $-Z$ axis. The explicit spinor solutions in the Pauli-Dirac representation are given for a particle of mass \(m\), energy \(E\), and momentum along or opposite to the \(z\)-axis. For helicity \(\lambda = \pm1/2\) states, the baryon spinors are:
\begin{equation}
u(p_z, \pm1/2) = \begin{pmatrix} \sqrt{E+m} \, \chi_\pm \\ \pm\sqrt{E-m} \, \chi_\pm \end{pmatrix}, \quad u(-p_z, \pm1/2) = \begin{pmatrix} \sqrt{E+m} \, \chi_\mp \\ \pm\sqrt{E-m} \, \chi_\mp \end{pmatrix},
\end{equation}
and the antibaryon spinors are:
\begin{equation}
v(p_z,\pm1/2) = \begin{pmatrix} \sqrt{E-m} \, \chi_\mp \\ \mp\sqrt{E+m} \, \chi_\mp \end{pmatrix}, \quad v(-p_z, \pm1/2) = \begin{pmatrix} -\sqrt{E-m} \, \chi_\pm \\ \pm\sqrt{E+m} \, \chi_\pm \end{pmatrix},
\end{equation}
with two-component basis spinors \(\chi_+ = (1,0)^T\) and \(\chi_- = (0,1)^T\).

Using these equations, one gets

\begin{align}
\bar{u}(P_q,1/2) v(-P_q,1/2)&=-2P_q, \nonumber\\
\bar{u}(P_q,1/2) \slashed{\epsilon} (0)v(-P_q,1/2)&=-2 m_q,\nonumber \\
\bar{u}(P_q,1/2) \slashed{\epsilon} (1)v(-P_q,-1/2)&=-2\sqrt{2}E_q, 
\label{Dirac}
\end{align}
where $P_q$ is the momentum magnitude for the quark $q$ with mass $m_q$ and energy $E_q$. The polarization vectors are taken as ${\epsilon }{\left( \pm 1 \right) } =  \mp  \frac{1}{\sqrt{2}}\left( {0,1, \pm  i,0}\right)^T$ and ${\epsilon }{\left( 0\right) } = \left( {0,0,0,1}\right)^T$.

Then the helicity amplitude for $\chi_{c2} \to  B \bar{B}$ are calculated as follows:
\begin{align}\label{chic2amp}
b_{+,+}(N \bar{N}) =& -4 \sqrt{\frac{2}{3}} \, P_q \, ( \frac{10}{3}m_q^2 -\frac{8}{9} E_q^2 ), \nonumber\\
b_{+,-}(N \bar{N}) =& -\frac{80}{9} \, E_q \, m_q \, P_q , \nonumber\\
b_{+,+}(\Lambda \bar{\Lambda}) =& -8 \sqrt{\frac{2}{3}} \, P_q \, \big( 2 m_q m_s - P_q P_s \big), \nonumber\\
b_{+,-}(\Lambda \bar{\Lambda}) =& -16 \, E_s \, m_q \, P_q ,\nonumber\\
b_{+,+}(\Sigma \bar{\Sigma}) =& -\frac{8}{3} \sqrt{\frac{2}{3}} \,
( 2(E_q^2+3 m_q^2) 
+ ( 12 m_s m_q -8 E_s E_q) P_q ), \nonumber\\
b_{+,-}(\Sigma \bar{\Sigma}) =& \frac{32}{3} \,
( 2 E_q ( m_q P_s + m_s P_q ) -  E_s m_q P_q ), \nonumber\\
b_{+,+}(\Xi \bar{\Xi}) =& -\frac{8}{3} \sqrt{\frac{2}{3}} \,
( E_s^2 P_q - 4 E_q E_s P_s  
+3 m_s ( P_q m_s + 2 m_q P_s ) ), \nonumber\\
b_{+,-}(\Xi \bar{\Xi}) =& \frac{16}{3} \,
( - E_q m_s P_s + 2 E_s ( P_q m_s + m_q P_s ) ).
\end{align}
Here $N \bar{N}$ denote proton-antiproton $p \bar{p}$ or neutron-antineutron $n \bar{n}$, $\Sigma \bar{\Sigma}$  denote $\Sigma^0 \bar{\Sigma}^0$ or $\Sigma^+ \bar{\Sigma}^-$, $\Xi \bar{\Xi}$ denote $\Xi^0 \bar{\Xi}^0$ or $\Xi^- \bar{\Xi}^+$ .
 The \( m_u, m_d, E_u, E_d \) are the masses and energies of  \( u \) and \( d \) quarks. Under the assumption \( m_u = m_d = m_q \) and \( E_u = E_d = E_q \), the light-quark momentum is \( P_q = \sqrt{E_q^2 - m_q^2} \). Correspondingly, for the strange quark (mass \( m_s \), energy \( E_s \)), we have \( P_s = \sqrt{E_s^2 - m_s^2}\).

In the numerical evaluation of helicity amplitudes, the light-quark mass is set to \( m_q = m_p / 3 = 0.31 \) GeV~\cite{mass:2000qj}, and the strange-quark mass is determined via \( m_s = (m_\Lambda - m_p) + m_q \) GeV, where \( m_\Lambda \) denotes the \(\Lambda\)-baryon mass, yielding \( m_s = 0.49 \) GeV. Quark momenta are obtained as \( P_q = m_q P_B / M \) and \( P_s = m_s P_B / M \), with \( M = m_u + m_d + m_s \) being the total quark mass and \( P_B \) the baryon momentum. We estimate the corresponding uncertainty by evaluating our results at \( m_s = 0.43 \) GeV and 0.53 GeV, and quote the larger difference as the uncertainty. Numerical results for the decays \( \chi_{c2} \to B\bar{B} \) are listed in Tab.~\ref{tab:helamp}. Due to the adoption of the \(m_u = m_d\) assumption, the processes for the $\Xi^-$ and $\Xi^0$  are treated unified. The calculated angular distribution parameters agree with the experimental data within one standard deviation. 
\begin{table*}[htbp]
  \centering
  \caption{Numerical helicity amplitudes and angular distribution parameters for the \(\chi_{c2}\) decay into octet baryonic pairs . $\alpha$ is the numerical estimation, and $\alpha$(Exp) is measured values.}
  \label{tab:helamp}
  \begin{tabular}{l  c c c c }
   \hline\hline
    Channel &$ b_{+-}$ &$ b_{++}$ & $\alpha$& $\alpha$(Exp) \\
    \hline
    $p \bar{p}$&$-0.816\pm 0.001$&$-0.016\pm 0.001$ &$-0.333\pm 0.001$&$-0.26\pm0.17$ \cite{BESIII:chicjpp} \\\hline
     $n \bar{n}$&$-0.816\pm 0.001$&$-0.016\pm 0.001$ &$-0.333\pm 0.001$& -\\\hline
   $\Lambda\bar{\Lambda}$&$-1.488\pm 0.191$ &$-0.177\pm 0.016$& $-0.337\pm0.001$ &  \(-0.211 \pm 0.100 \pm 0.050\) \cite{BESIII:2025chicj2LL}  \\\hline
$\Sigma^0\bar{\Sigma}^0$&$2.762\pm0.221$&$-0.438\pm 0.153$& $-0.341\pm0.007$&-\\\hline
$\Sigma^+\bar{\Sigma}^-$&$2.762\pm0.221$&$-0.438\pm 0.153$& $-0.341\pm0.007$&-\\\hline
      $\Xi^{-}\bar{\Xi}^{+}$&$ 1.824\pm0.409$&$-0.461\pm0.121$&$-0.351\pm0.006$
           &  $-0.34 \pm 0.18 \pm 0.30$  \cite{BESIII:2022chicjXX} \\\hline
$\Xi^{0}\bar{\Xi}^{0}$&$1.824\pm0.409$&$-0.461\pm0.121$&$-0.351\pm0.006$
      &$-0.65 \pm 0.31 \pm 0.22$  \cite{BESIII:2022chicjXX} \\
   \hline \hline
  \end{tabular}
\end{table*}

\section{polarization observable}
\subsection{Angular distribution}
\begin{figure}[htbp]
\begin{center}
\begin{minipage}[t]{0.35\linewidth}
\includegraphics[width=1\textwidth]{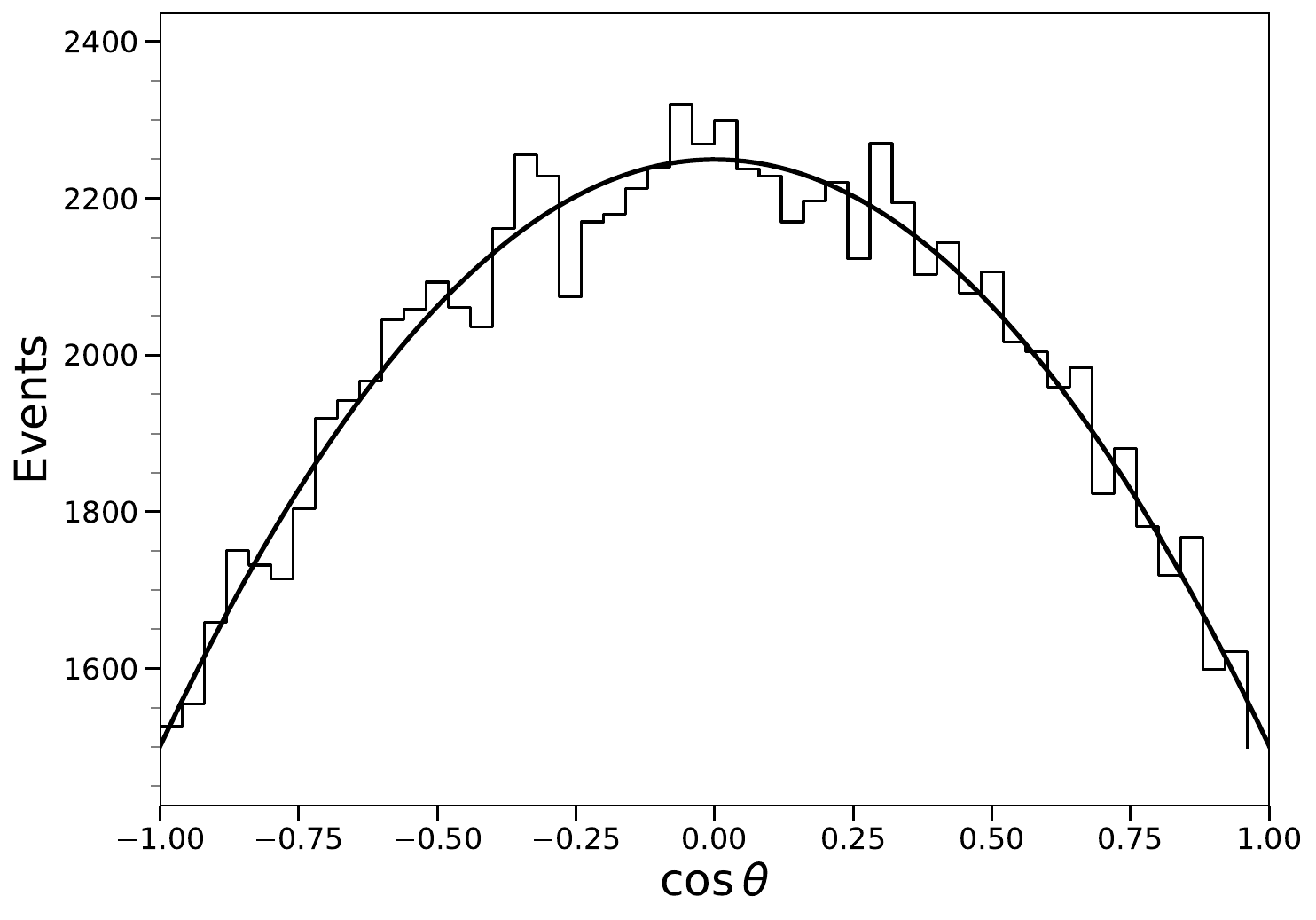}
\put(-90,60){ (a)}
\end{minipage}
\begin{minipage}[t]{0.35\linewidth}
\includegraphics[width=1\textwidth]{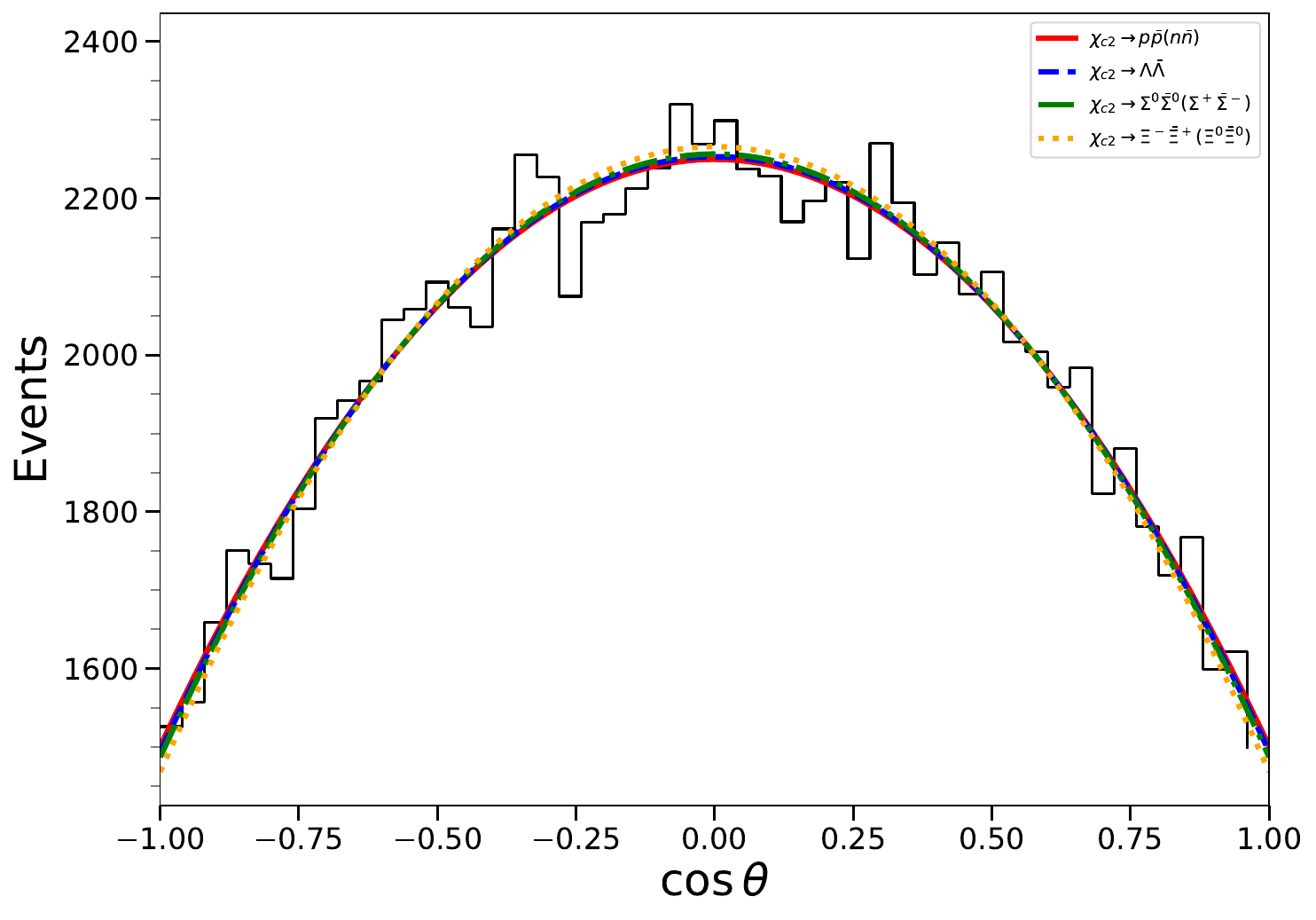}
\put(-90,60){ (b)}
\end{minipage}
\end{center}
\caption{Angular distributions of octet baryon  in $\chi_{c1}$ (a) and $\chi_{c2}$ (b) decays. 
where histograms are MC events, and the curves are the expected distribution of $1 - \alpha \cos^2\theta_1$ with $\alpha = -1/3$ (a) and $\alpha = 0.333,0.337,0.341,0.351$ for $p(n),\Lambda,\Sigma^0(\Sigma^+),\Xi^0(\Xi^+)$ (b).}
\label{chic12_angular}
\end{figure}

In the decay processes $\chi_{cJ}$ to spin-1/2 baryon antibaryon pair, the polar angular distribution of the final-state baryon in the helicity frame provides a direct probe into the parent particle's polarization and the underlying helicity amplitudes. The angular distribution can be generally expressed as $I(\theta_1) \propto 1 + \alpha\cos^2\theta_1$. For the scalar state $\chi_{c0}$, the parameter $\alpha$ is predicted to be zero, resulting in a flat, isotropic distribution, a direct consequence of its $J=0$ nature, which forbids any spin alignment. In contrast, for the axial-vector $\chi_{c1}$, the decay into any octet baryon-antibaryon pair is governed by a helicity selection rule enforced by charge conjugation symmetry. This rule strictly constrains the helicity amplitudes, leading to a model-independent prediction of $\alpha = -1/3$ for all such channels. 
Based on the theoretical formula given in Eq.~(\ref{chic1:alpha}), we generate an ensemble of MC events   and obtain the distribution for $\chi_{c1}\to B\bar{B}$ shown in Fig.~\ref{chic12_angular}(a).
This remarkable universality makes the $\chi_{c1}$ decay a clean testbed for fundamental symmetries in QCD. For instance, the recent measured value by BESIII for $\chi_{c1} \to \Lambda\bar{\Lambda}$, $\alpha = -0.301 \pm 0.042 \pm 0.026$~\cite{BESIII:2025chicj2LL}, is in excellent agreement with this prediction, confirming the dominance of the helicity selection rule in this decay mode.

For the tensor state $\chi_{c2} \to B\bar{B}$, the situation is more complex. The angular distribution parameter $\alpha$ is also predicted to be negative, but its exact value is not universal; it depends on the relative magnitude and phase of two independent helicity amplitudes. These amplitudes are sensitive to the detailed dynamics of the decay, such as the wave function overlap and the spin structure of the final-state baryons, and thus can be calculated within specific frameworks like the quark model.
We use the angular distribution formula in Eq.~(\ref{chic2 angur})
and the parameter $\alpha$ fixed from the helicity amplitudes in Table~\ref{tab:helamp}, to generate an ensemble of MC events 
using an acceptance-rejection method for the $\chi_{c2}$ decays.
Figure~\ref{chic12_angular}(b) shows the angular distribution for $\chi_{c2}\to B\bar{B}$,  histogram is filled with the MC events, and the comparison with the predicted angular distribution (curve).

The angular distributions in different baryon channels ($p\bar{p}$, $\Lambda\bar{\Lambda}$, $\Sigma^0\bar{\Sigma}^0$($\Sigma^+\bar{\Sigma}^-$), $\Xi^-\bar{\Xi}^+$($\Xi^{0}\bar{\Xi}^{0}$)) exhibit subtle yet discernible differences in the Fig.~\ref{chic12_angular}, reflecting the dependence of the decay amplitudes on the flavor structure of the final-state baryons.
Current experimental measurements for $\chi_{c2}$ decays into $p\bar{p}$, $\Lambda\bar{\Lambda}$ and $\Xi^-\bar{\Xi}^+$ ($\Xi^{0}\bar{\Xi}^{0}$) all yield negative $\alpha$ values~\cite{BESIII:chicjpp,BESIII:2025chicj2LL,BESIII:2022chicjXX}, and the results are consistent with quark-model calculations within one standard deviation. To more stringently test these theoretical predictions and decode the intricate decay mechanisms of $\chi_{cJ}$ states, future high-statistics experiments are crucial. Improving the precision of existing measurements and performing the first measurement for the $\chi_{c2} \to \Sigma^{0}\bar{\Sigma}^{0}$ channel will provide essential data. These precise determinations of $\alpha$ across the baryon octet will serve as critical benchmarks, allowing us to distinguish between different model assumptions and ultimately achieve a deeper understanding of the interplay between perturbative and non-perturbative QCD in charmonium decays.

\subsection{Baryonic polarization and spin correlation}
In the decay $\chi_{c0} \to B\bar{B}$, the parent particle's spin of zero forbids any net polarization of the final-state baryons along any axis ($x$, $y$, or $z$). However, the spins of the baryon and antibaryon remain strongly quantum-correlated. These non-classical spin correlations, a signature of entanglement, are not directly visible in the primary decay products but can be observed through momentum-angle correlations among the decay particles from the subsequent weak decays of both the baryon and antibaryon.

\begin{figure}[htbp]
\centering
\includegraphics[width=0.38\textwidth]{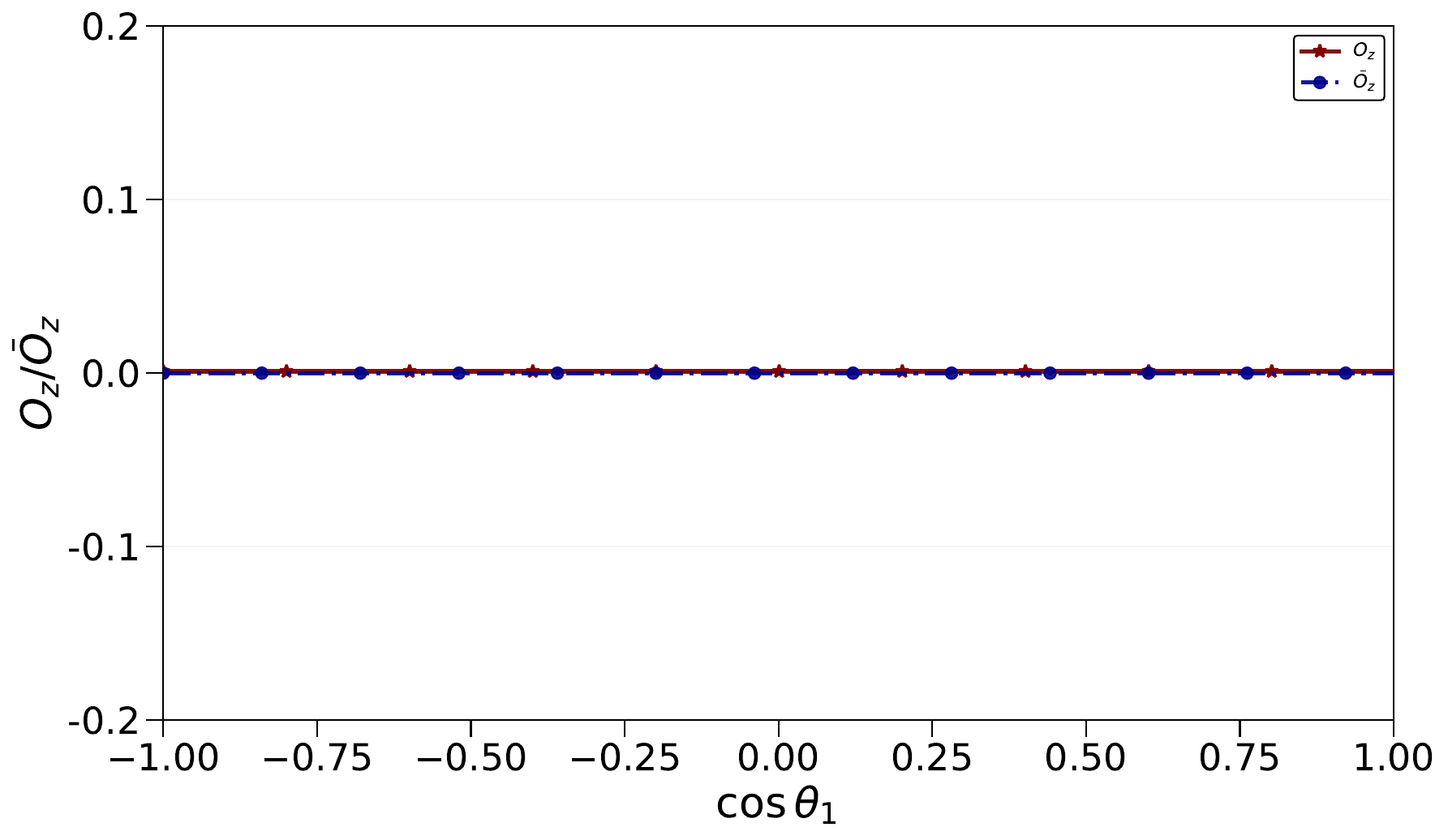}
\caption{ The  dependence of polarization $O_z, \bar{O}_z$ 
versus $\cos\theta_1$ for baryon and antibaryon  in $\chi_{c1} \to B\bar{B}$($B$:octet baryon). }
\label{chic1_Pz}
\end{figure}

\begin{figure}[htbp]
\centering
\includegraphics[width=0.38\textwidth]{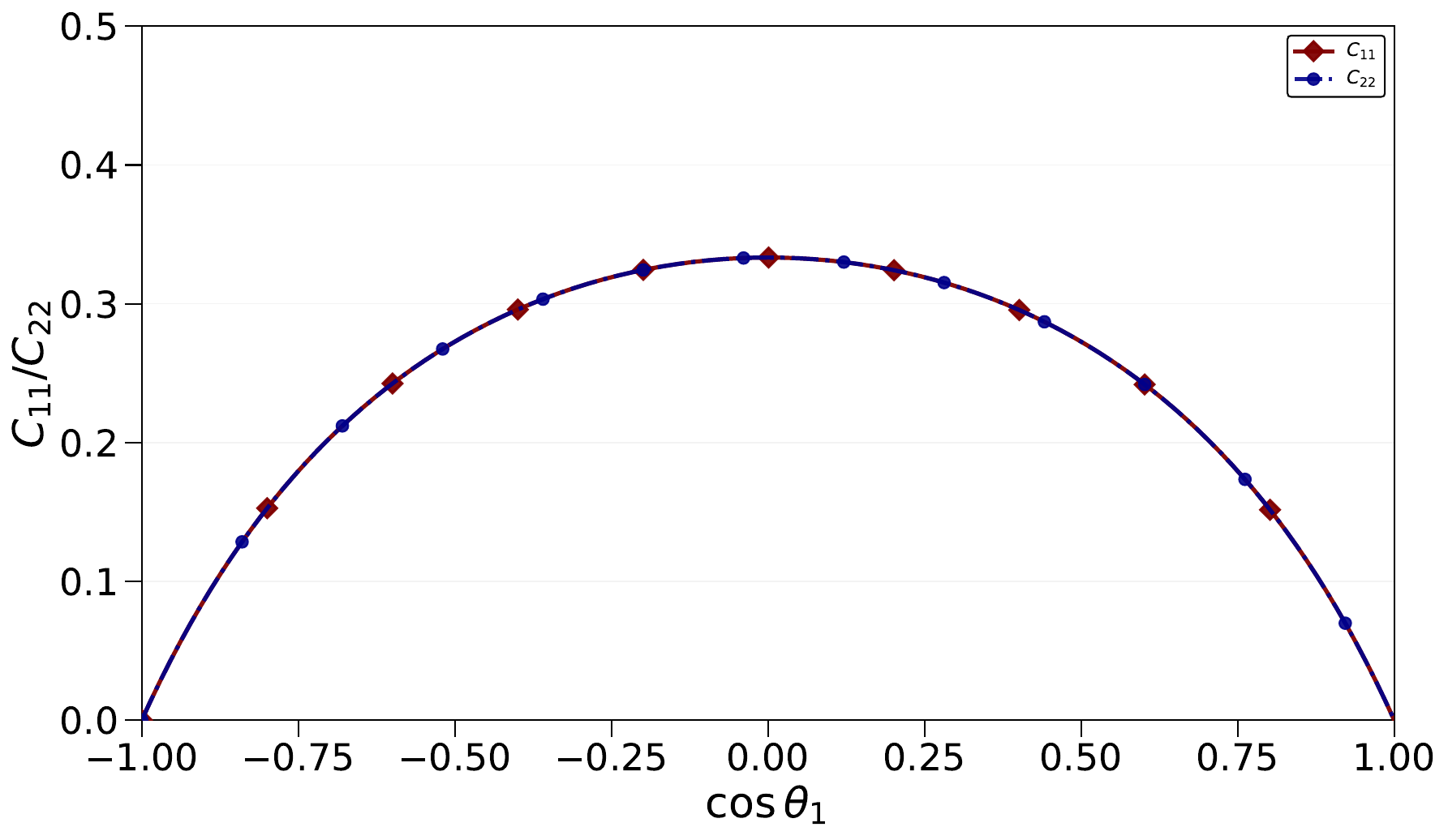}
\caption{ The dependence of spin correlation matrix elements $C_{ij}$ versus $\cos\theta_1$ in $\chi_{c1} \to B\bar{B}$($B$:octet baryon). }
\label{chic1_Cij}
\end{figure}

For the process $\chi_{c1} \to B\bar{B}$, the initial longitudinal polarization of the colliding beams is transferred to the $\chi_{c1}$ and manifests as a longitudinal polarization of the final-state baryon and antibaryon. This polarization is directly measurable. Furthermore, the spin correlations between the baryon pair are encoded in their joint angular distribution, specifically in the dependence on the polar angle $\theta_1$, allowing for a complete analysis of the spin state from the primary decay vertex.

Figure~\ref{chic1_Pz} presents the longitudinal polarization components \(O_z\) and \(\bar{O}_z\) of the baryon and antibaryon as functions of \(\cos\theta_1\) in these decays . In all channels, \(O_z\)(\(\bar{O}_z\)) is found to be consistent with zero, in agreement with the suppression of longitudinal polarization expected from helicity conservation in pseudoscalar‑meson decays. Figure.~\ref{chic1_Cij} shows that the distributions of the spin correlation matrix elements in these decays are consistent across different baryon channels, indicating that the spin correlation pattern in $\chi_{c1}$ decays is primarily governed by the symmetry of the initial state and exhibits negligible dependence on the flavor structure of the final-state baryons.

The decay $\chi_{c2} \to B\bar{B}$ yields a different polarization pattern. Here, the baryon and antibaryon possess only a non-zero transverse polarization along the $y$-axis (perpendicular to the production plane). The magnitude of this transverse polarization depends on the ratio and relative phase between the two independent helicity amplitudes governing the decay. As in the $\chi_{c1}$ case, the spin correlations between the pair are also embedded in and can be extracted from the polar angular distribution $I(\cos\theta_1)$.

\begin{figure}[htbp]
\centering
\includegraphics[width=0.35\textwidth]{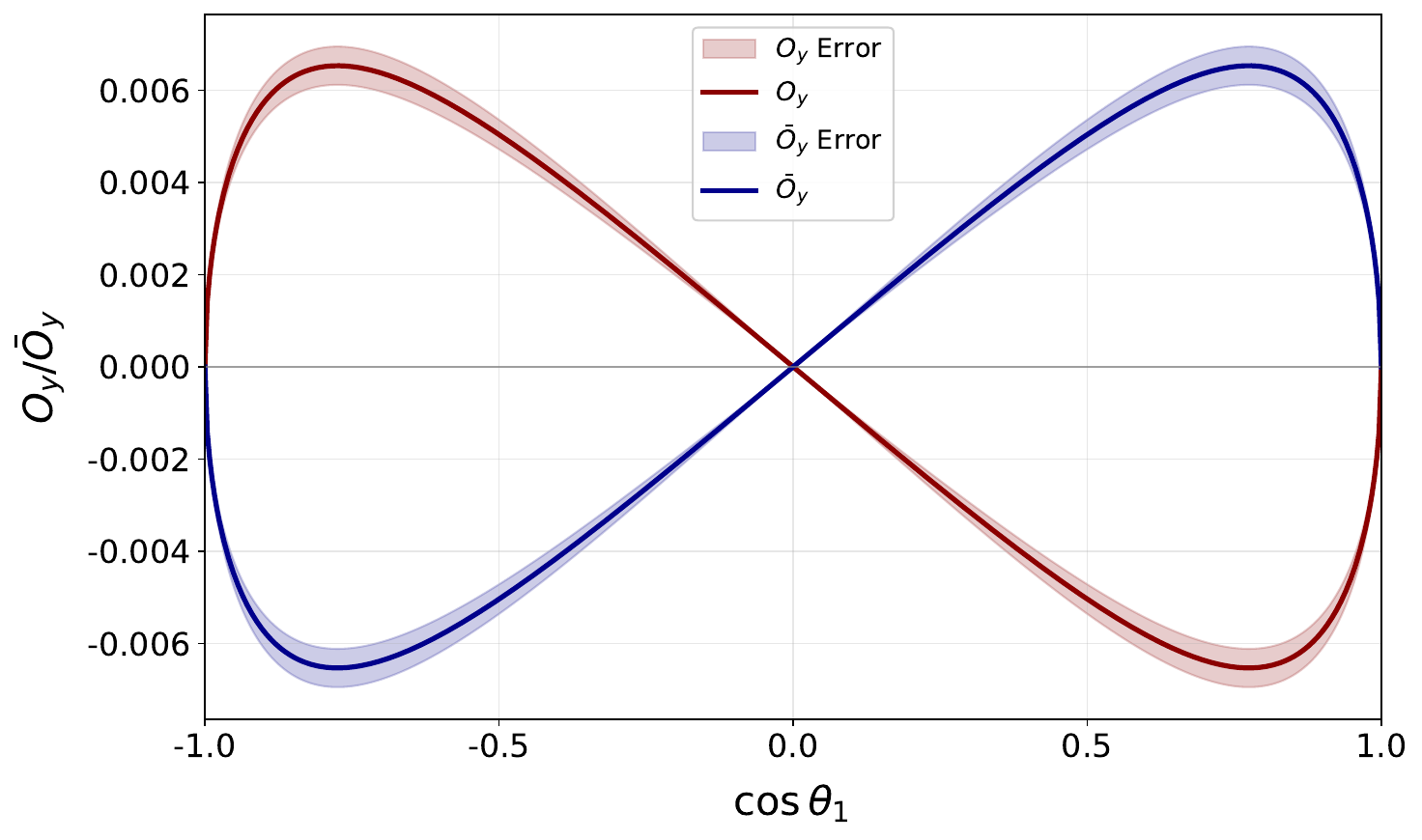}
\includegraphics[width=0.35\textwidth]{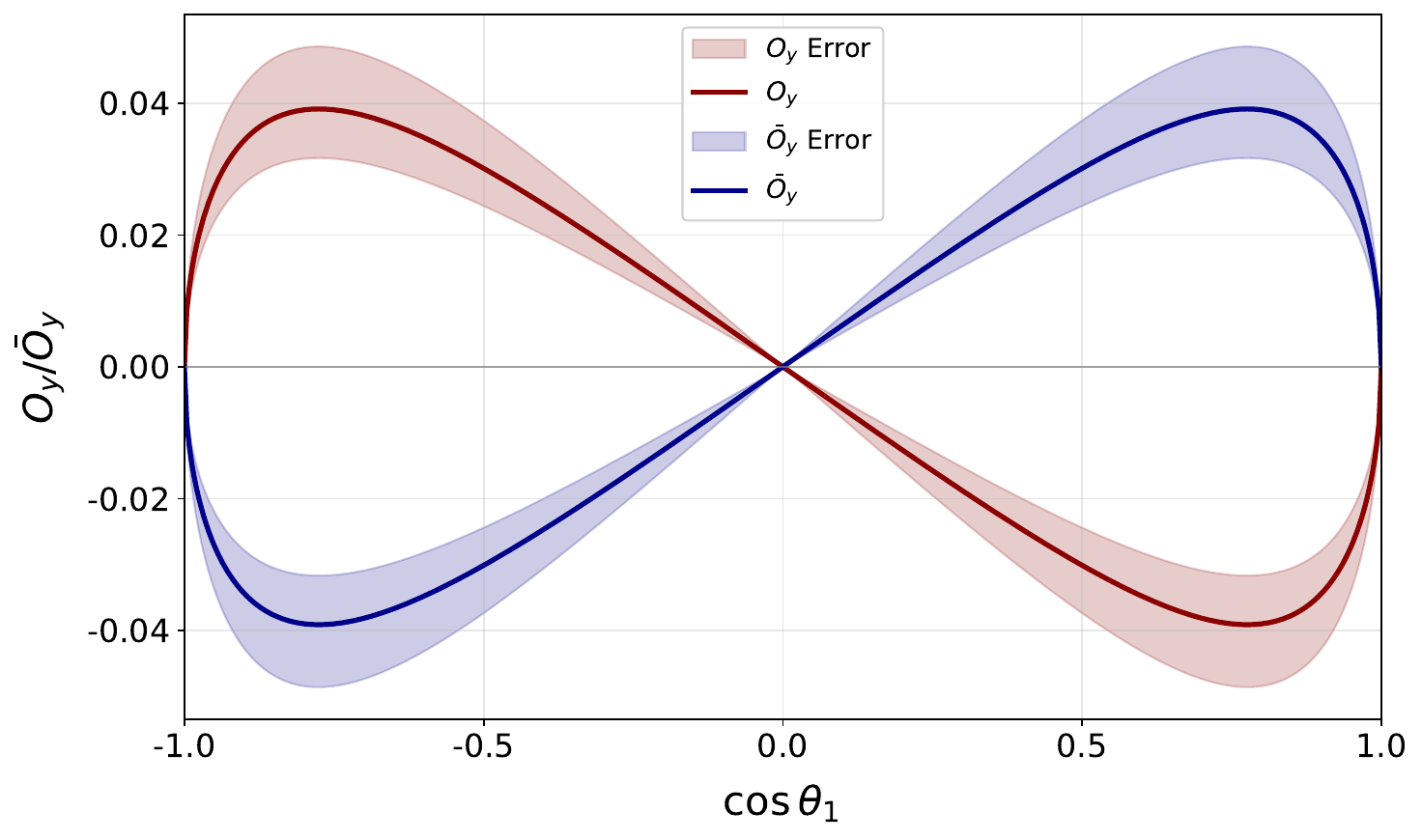}  
\includegraphics[width=0.35\textwidth]{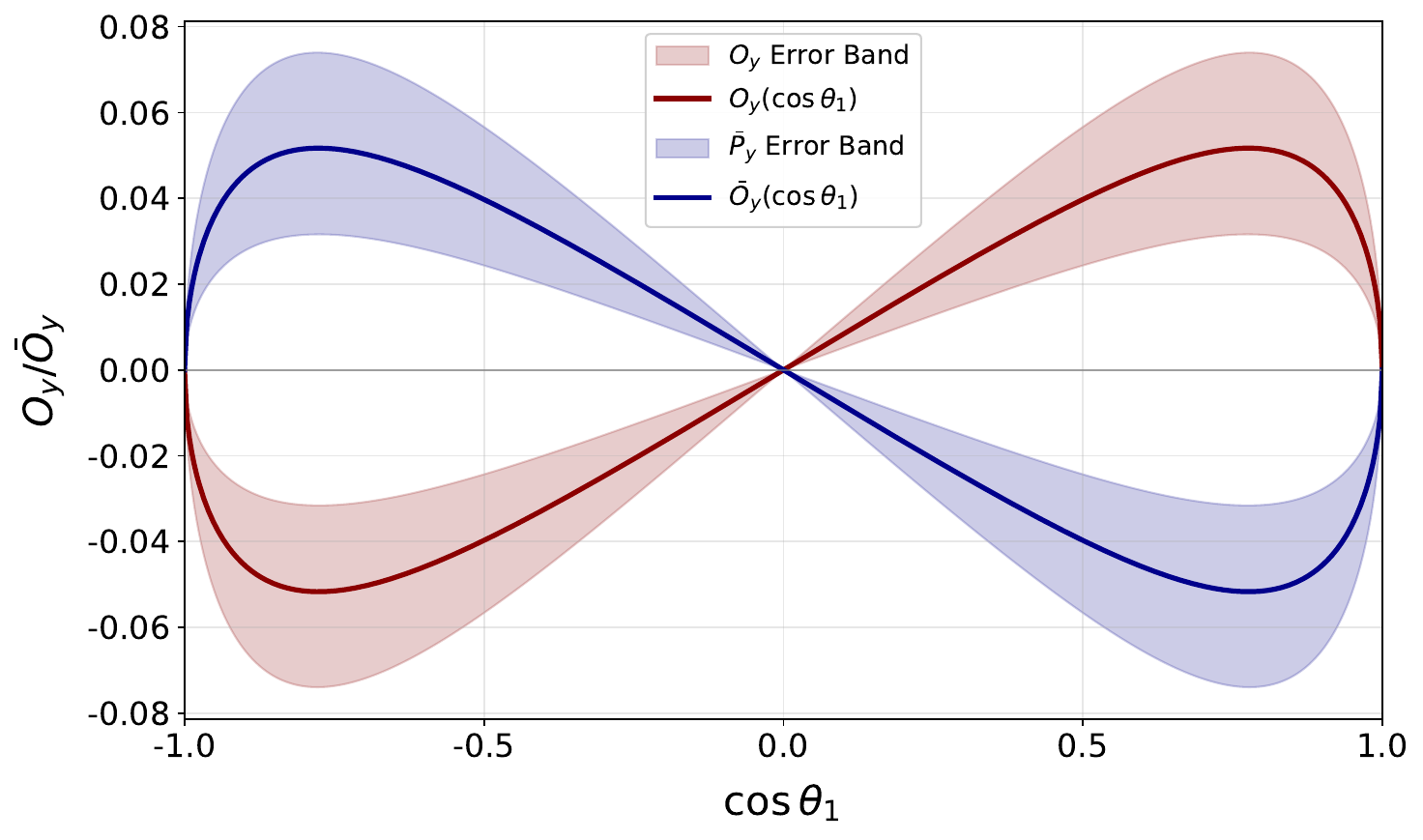}  
\includegraphics[width=0.35\textwidth]{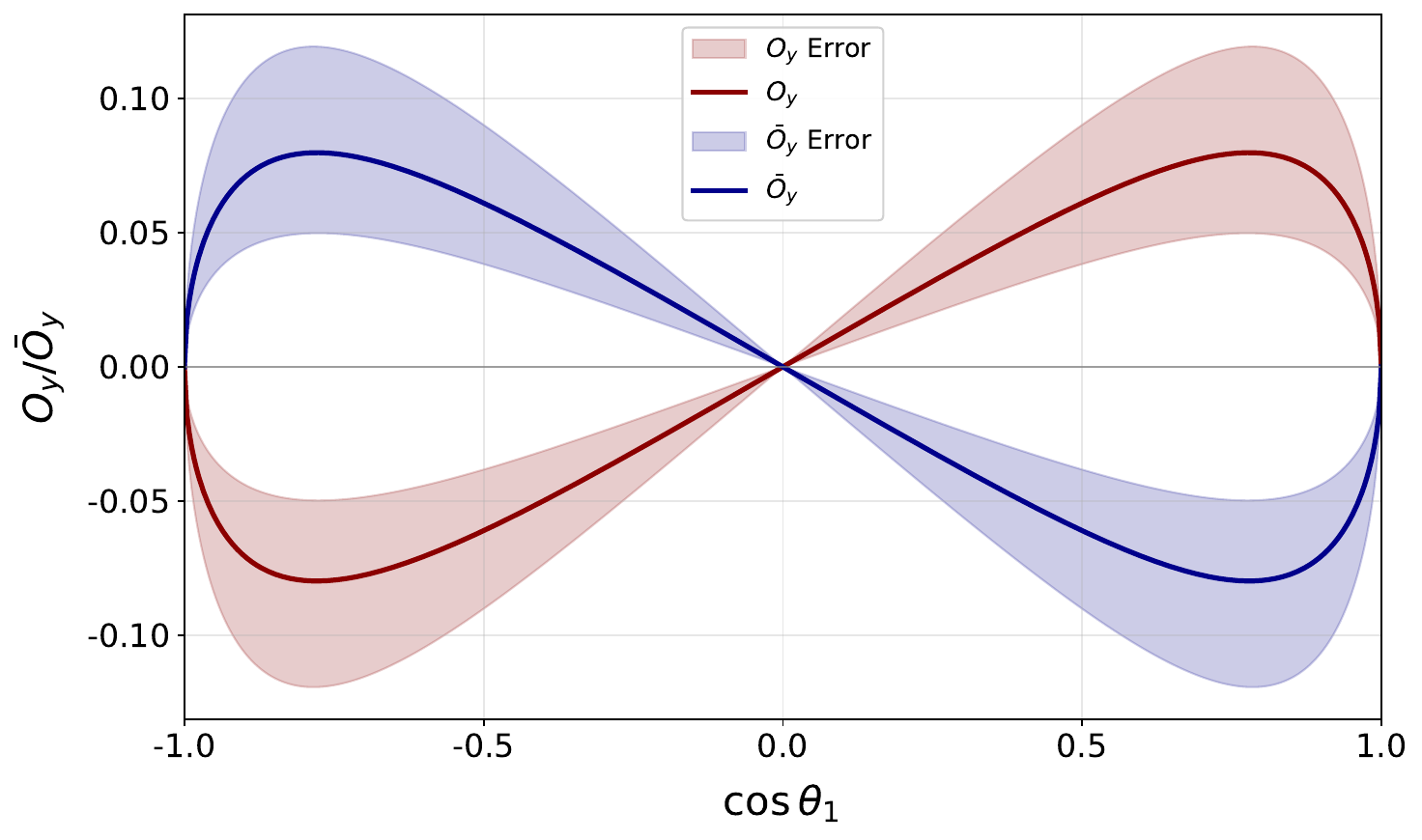}
\caption{The polarization components $O_y$ and $\bar{O}_{y}$ of baryons and antibaryons from $\chi_{c2}$ decays are shown as follows: upper-left for $p\bar{p}$ ($n\bar{n}$), upper-right for $\Lambda\bar{\Lambda}$, bottom-left for $\Sigma^0\bar{\Sigma}^0$ ($\Sigma^+\bar{\Sigma}^-$), and bottom-right for $\Xi^-\bar{\Xi}^+$ ($\Xi^0\bar{\Xi}^0$). The parameters $b_{+-}$ and $b_{++}$ are set to the numerical values in Table I. The solid line uses their central values, and the shaded band denotes the uncertainties.}
    \label{chic2_Py}
\end{figure}

\begin{figure}[htbp]
\centering
\includegraphics[width=0.9\textwidth]{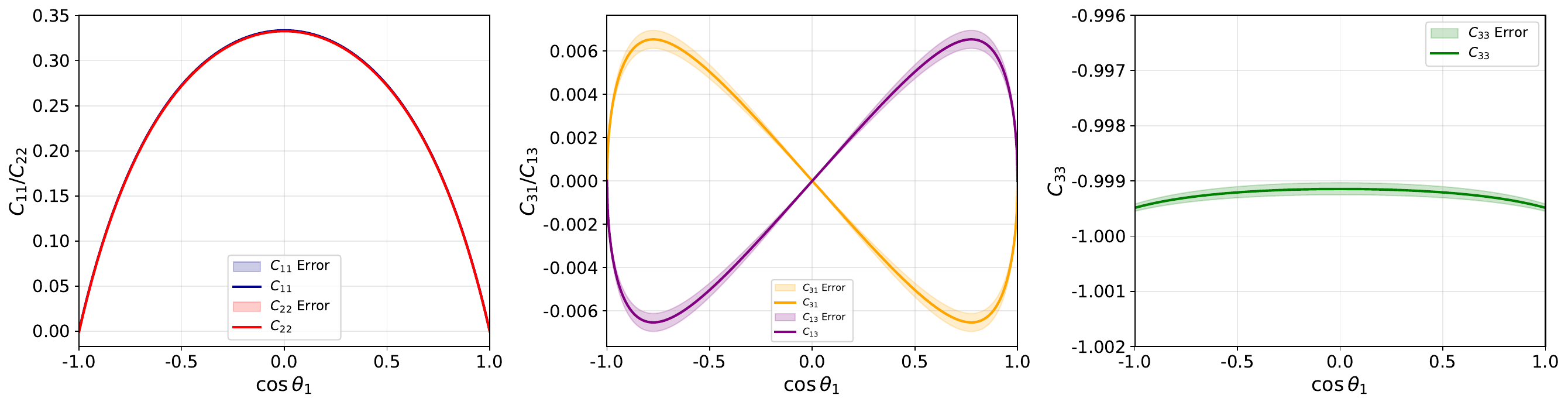}
\includegraphics[width=0.9\textwidth]{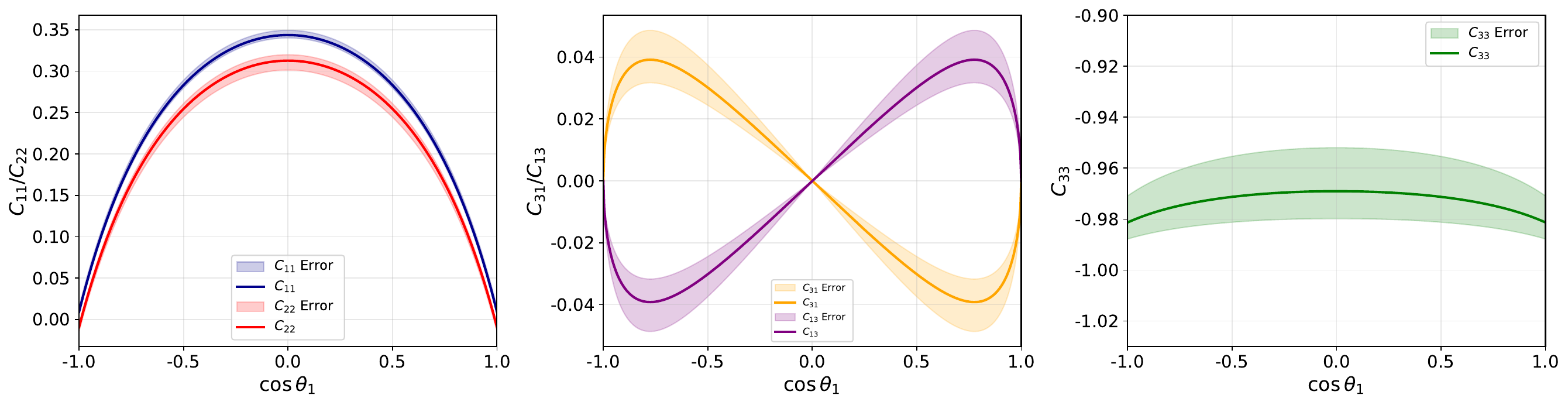}
\includegraphics[width=0.9\textwidth]{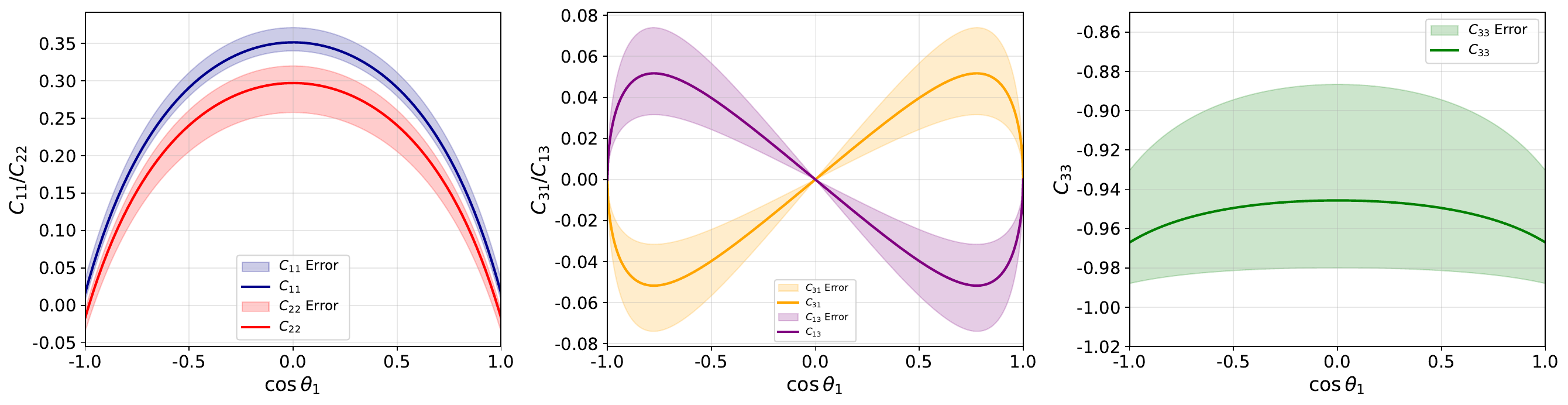}
\includegraphics[width=0.9\textwidth]{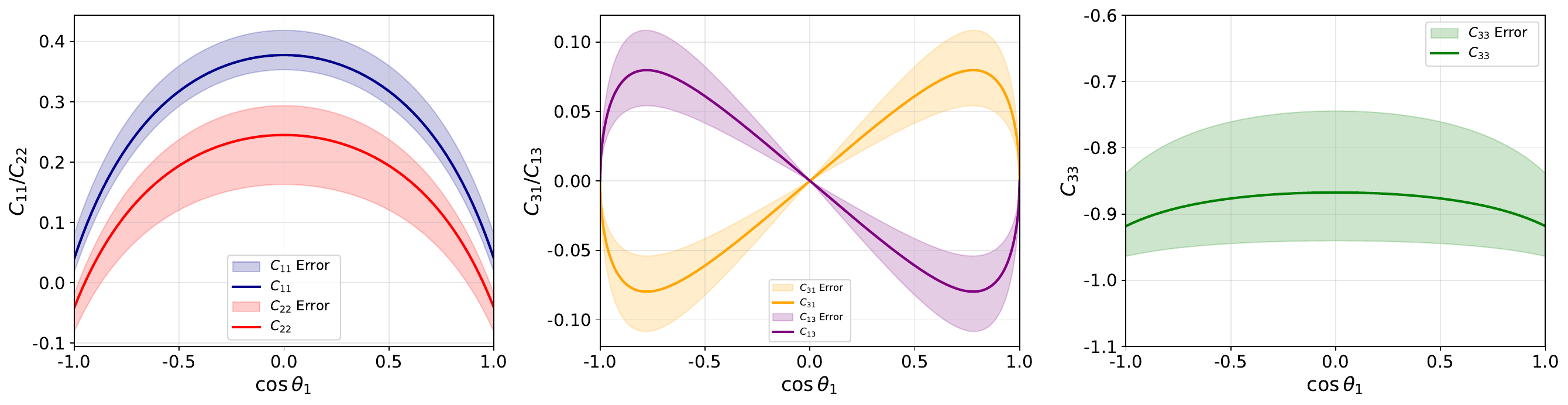}
\caption{The spin correlation matrix elements $C_{ij}$. The first row corresponds to $\chi_{c2} \to p\bar{p}$, the second to $\chi_{c2} \to \Lambda\bar{\Lambda}$, the third to $\chi_{c2} \to \Sigma^0\bar{\Sigma}^0$, and the fourth to $\chi_{c2} \to \Xi^-\bar{\Xi}^+$ ($\Xi^0\bar{\Xi}^0$). The parameters $b_{+-}$ and $b_{++}$ are set to the numerical values in Table I. The solid line uses their central values, and the shaded band denotes the uncertainties.}
\label{spin-cor}
\end{figure}

Based on Eq.~(\ref{chic2_O_y}) to Eq.~(\ref{chic2_C_ij}) and the helicity amplitude values derived from the naive quark model in Table.~\ref{tab:helamp}, the transverse polarization  and the spin correlation matrix for the process$\chi_{c2}\to B\bar{B}$ can be obtained under the assumption of quark masses and energies. The phase angle \(\Delta\) can only be determined by fitting to experimental data, such as the value \(\Delta = -0.37\pm0.16\) provided by BESIII for the \(\chi_{c2} \to \Lambda \bar{\Lambda}\) decay~\cite{BESIII:2025chicj2LL}.  When predicting the polarization vector and spin correlation tensor, we set \(\Delta = \pi/2\) and  \(\Delta = 0\) in our simulations.
As shown in Fig.~\ref{chic2_Py}, a non‑zero transverse polarization  $O_y$ is observed in all decay channels, with its magnitude following a flavor‑dependent pattern. 
Figure~\ref{spin-cor} shows the measurements of the spin correlation matrix \(C_{ij}\) in \(\chi_{c2} \to B\bar{B}\) decays (\(B = p,\Lambda,\Sigma^0(\Sigma^+),\Xi^-(\Xi^0)\)). A strong longitudinal alignment (\(C_{33} \approx 1\)) and universal transverse anisotropy (\(C_{11} > C_{22}\)) across all channels. The effect is most pronounced for \(\Xi^-\bar{\Xi}^+(\Xi^0\bar{\Xi}^0)\), suggesting that interference among helicity amplitudes is modulated by baryon spin structure. These results link charmonium decay dynamics to flavor-dependent baryon structure.

\section{Summary and outlook}

In this work, we have performed a comprehensive polarization analysis of $\chi_{cJ}$ ($J=0,1,2$) decays into spin-1/2 baryon-antibaryon pairs, with a focus on the novel scenario where the $\chi_{cJ}$ states are produced via polarized electron-positron collisions. Starting from the spin density matrix formalism, we systematically derived the polarization transfer from the initial beam to the intermediate $\psi(2S)$ and $\chi_{cJ}$ resonances, and finally to the baryonic final states. For the subsequent decay $\chi_{cJ} \to B\bar{B}$, we calculated the complete set of helicity amplitudes within a constituent quark model framework, incorporating the full SU(6) spin-flavor wave functions for octet baryons.

Our key findings can be summarized as follows. First, the angular distribution parameter $\alpha$ for $\chi_{c1} \to B\bar{B}$ is confirmed to be universally $-1/3$, a robust prediction arising from the helicity selection rule imposed by charge-conjugation symmetry. Second, in the case of $\chi_{c2}$ decays, the $\alpha$ parameter, the transverse polarization along the $y$-axis, and the spin correlation tensor all depend on the ratio and relative phase of two independent helicity amplitudes. Our calculated results are consistent with available experimental measurements within the quoted uncertainties. Finally, under a polarized beam scenario, the initial longitudinal beam polarization $P_z$ modifies the spin density matrices of $\chi_{c1}$ and $\chi_{c2}$, which induces observable effects in the final-state baryon polarizations and spin correlations. By contrast, the decay of $\chi_{c0}$ remains insensitive to beam polarization.

Looking forward, our work provides a theoretical toolkit for future experiments utilizing polarized beams, such as the proposed STCF. Measurements of the detailed angular distributions and spin correlations with polarized $e^+e^-$ collisions will offer unprecedented precision to test the quark model calculations, scrutinize the helicity selection rules, and probe potential contributions beyond the dominant E1 transition. Furthermore, the predicted strong spin entanglement in $\chi_{c0}$ decays and the polarization-dependent entanglement witness in $\chi_{c1,2}$ decays open a promising avenue to explore quantum entanglement in high-energy physics, potentially using these processes as a sensitive probe for new physics beyond the Standard Model.

\section{ ACKNOWLEDGEMENTS}
The work is partly supported  by the National Natural Science Foundation of China (NSFC) under Grants No. 12575112 and No. 12465015.

\newpage
\appendix
\section{spin wave function of $\chi_{c2}$}
The spin wave functions of $\chi_{c2}$ in its rest frame are taken as 
 \begin{equation}
 \begin{split}
 \label{epsilon}
 \epsilon_{\mu\nu}(0)&=\frac{1}{\sqrt{6}}(\epsilon_{\mu}(1)\epsilon_{\nu}(-1)+2\epsilon_{\mu}(0)\epsilon_{\nu}(0)+\epsilon_{\mu}(-1)\epsilon_{\nu}(1)),\\
 \epsilon_{\mu\nu}(\pm 1)&=\frac{1}{\sqrt{2}}(\epsilon_{\mu}(\pm 1)\epsilon_{\nu}(0)+\epsilon_{\mu}(0)\epsilon_{\nu}(\pm 1)),\\
 \epsilon_{\mu\nu}(\pm 2)&=\epsilon_{\mu}(\pm 1)\epsilon_{\nu}(\pm 1),
\end{split}
\end{equation}
here $\epsilon(\pm 1 )  = \mp  \frac{1}{\sqrt{2}}( 0,1, \pm i,0)^T$ and $\epsilon( 0) = ( 0,0,0,1)^T$.

\section{Baryon wave function}

In the naive quark model, the wave functions are constructed as follows. For the proton and neutron, by neglecting the mass difference between the \(u\) and \(d\) quarks, its flavor-spin wave function can be explicitly expressed within the SU(6) group representation. Within this framework, the flavor wave functions of the proton and neutron are given by:
\begin{equation}
\begin{split}
&\phi ^{\rho}_p = \frac{1}{\sqrt{2}}(udu-duu), ~~\phi ^{\rho}_n = \frac{1}{\sqrt{2}}(udd-dud), \\
&\phi^{\lambda}_p =- \frac{1}{\sqrt{6}}(duu+udu-2uud),~~\phi^{\lambda}_n = \frac{1}{\sqrt{6}}(udd+dud-2ddu)
\end{split}
\end{equation}
where $\phi^{\rho}_{p,n}$ and $\phi ^{\lambda}_{p,n}$ denote the  antisymmetric and symmetric flavor wave functions of the proton and neutron, respectively.
The overall flavor wave function exhibits mixed symmetry.
The spin wave function of the spin-$1/2$ state with spin projection $\lambda=1/2$ is given as follows:
\begin{equation}
\label{spin}
\begin{split}
\chi_{\frac{1}{2}}^{\rho} &= \frac{1}{\sqrt{2}}\left( {\left| { \uparrow   \downarrow   \uparrow  \rangle  - }\right|  \downarrow   \uparrow   \uparrow  \rangle }\right), \\
\chi _{\frac{1}{2}}^{\lambda} &= -\frac{1}{\sqrt{6}}\left( |\downarrow   \uparrow  \uparrow  \rangle  + |  \uparrow \downarrow  \uparrow  \rangle -2|  \uparrow  \uparrow \downarrow   \rangle \right).
\end{split}
\end{equation}
For proton and neutron, the total spin-flavor wave function  reads
$\Psi_{p} = \frac{1}{\sqrt{2}}(\chi^\rho \phi^\rho + \chi^\lambda \phi^\lambda)$.

For octet baryons($\Lambda, \Sigma, \Xi$), the flavor wave functions for baryons are taken as
\begin{equation}
\begin{aligned}
\phi_{\Lambda} &= \frac{1}{\sqrt{2}}(uds-dus), \\
\phi_{\Sigma^0} &= \frac{1}{\sqrt{2}}(uds+dus), \\
\phi_{\Sigma^+} &= uus, \\
\phi_{\Xi^{0,-}}&= ssu,ssd.
\end{aligned}
\end{equation}

The spin wave function for the spin $1/2$ states with $z$ -projection $\lambda  = 1/2$ read
\begin{equation}
\begin{aligned}
\chi_{\Lambda} &= \frac{1}{\sqrt{2}}( {\left| { \uparrow   \downarrow   \uparrow  \rangle  - }\right|  \downarrow   \uparrow   \uparrow  \rangle }),\\
\chi_{\Sigma^{0,+}} &= \frac{1}{\sqrt{6}}( |\uparrow \downarrow  \uparrow   \rangle  + |  \downarrow\uparrow \uparrow \rangle -2|   \uparrow \uparrow \downarrow   \rangle ),\\
\chi_{\Xi^{0,-}} &= \frac{1}{\sqrt{6}}( |\uparrow \downarrow  \uparrow   \rangle  + |  \downarrow\uparrow \uparrow \rangle -2|   \uparrow \uparrow \downarrow   \rangle ).
\end{aligned}
\end{equation}
Thus the wave functions are constructed as $\Psi_Y=\phi_Y\chi_Y$ with $Y=\Lambda,\Sigma^0(\Sigma^+)$ and $\Xi^0$($\Xi^-)$.


\begin{thebibliography}{99}

\bibitem{Bolz:1997ez}
J.~Bolz, P.~Kroll and G.~A.~Schuler,
\href{doi:10.1007/s100520050174}{Eur. Phys. J. C \textbf{2}, 705--719 (1998)}


\bibitem{Ma:2001ri}
B.~Q.~Ma, I.~Schmidt, J.~Soffer and J.~J.~Yang,
\href{doi:10.1103/PhysRevD.65.034004}{Phys. Rev. D \textbf{65}, 034004 (2002)}

\bibitem{Liu:2010um}
X.~H.~Liu and Q.~Zhao,
\href{doi:10.1088/0954-3899/38/3/035007}{J. Phys. G \textbf{38}, 035007 (2011)}



\bibitem{BESIII:2018cnd}
M.~Ablikim \textit{et al.} (BESIII Collaboration),
\href{doi:10.1038/s41567-019-0494-8}{Nature Phys. \textbf{15}, 631--634 (2019)}

\bibitem{Perotti:2018wxm}
Elisabetta~Perotti, G{\"o}ran~F{\"a}ldt, Andrzej~Kupsc, Stefan~Leupold and Jiao Jiao~Song,
\href{doi:10.1103/PhysRevD.99.056008}{Phys. Rev. D \textbf{99}, 5, 056008 (2019)}

\bibitem{Moortgat-Pick:2005jsx}
G.~Moortgat Pick, T.~Abe \textit{et al.},
\href{doi:10.1016/j.physrep.2007.12.003}{Phys. Rept. \textbf{460}, 131--243 (2008)}

\bibitem{Zhang:2025oks}
Zhe~Zhang, Tianbo~Liu, Rong Gang~Ping, Jiao Jiao~Song and Weihua~Yang,
\href{doi:10.1103/mzkm-qw8w}{Phys. Rev. D \textbf{112}, 9, 096012 (2025)}

\bibitem{Bai2013chicj2BB}
M.~Ablikim \textit{et al.} (BESIII Collaboration),
\href{doi:10.1103/PhysRevD.87.032007}{Phys. Rev. D \textbf{87}, 3, 032007 (2013)}

\bibitem{BESIII:chicjpp}
M.~Ablikim \textit{et al.} (BESIII Collaboration),
\href{doi:10.1103/PhysRevD.88.112001}{Phys. Rev. D \textbf{88}, 11, 112001 (2013)}

\bibitem{Bai2003chicj2LL}
J.~Z.~Bai \textit{et al.} (BES Collaboration),
\href{doi:10.1103/PhysRevD.67.112001}{Phys. Rev. D \textbf{67}, 112001 (2003)}

\bibitem{BESIII:2025chicj2LL}
Medina~Ablikim \textit{et al.} (BESIII Collaboration),
\href{arXiv:2509.00289 [hep-ex]}{[arXiv:2509.00289 [hep-ex]]~(2025)} 



\bibitem{Bai2020chicj2SS}
M.~Ablikim \textit{et al.} (BESIII Collaboration),
\href{doi:10.1103/PhysRevD.101.092002}{Phys. Rev. D \textbf{101}, 092002 (2020)}

\bibitem{BESIII:2022chicjXX}
M.~Ablikim \textit{et al.} (BESIII Collaboration),
\href{doi:10.1007/JHEP06(2022)074}{JHEP \textbf{06}, 74 (2022)}


\bibitem{Horodecki:2009zz}
Ryszard~Horodecki, Pawel~Horodecki, Michal~Horodecki and Karol~Horodecki,
\href{doi:10.1103/RevModPhys.81.865}{Rev. Mod. Phys. \textbf{81}, 865--942 (2009)}

\bibitem{Adesso:2007tx}
Gerardo~Adesso and Fabrizio~Illuminati,
\href{doi:10.1088/1751-8113/40/28/S01}{J. Phys. A \textbf{40}, 7821--7880 (2007)}

\bibitem{Bernal:2024xhm}
Alexander~Bernal, Pawe{\l}~Caban and Jakub~Rembieli{\'n}ski,
\href{doi:10.1038/s41598-025-07747-3}{Sci. Rep. \textbf{15}, 1, 23410 (2025)}

\bibitem{Hong:2025drg}
P.~C.~Hong, R.~G.~Ping and W.~M.~Song,
\href{doi:10.1103/8bx4-gfc4}
{Phys. Rev. D \textbf{113}, no.7, 076009 (2026)}

\bibitem{Li:2026bkf}
C.~Li, X.~Cao, A.~Q.~Guo, C.~X.~Yu, H.~W.~Zhang and Z.~Zhang,
\href{arXiv:2602.10398 [hep-ph]}{arXiv:2602.10398 2026}.

\bibitem{ATLAS:2023fsd}
Georges~Aad \textit{et al.} (ATLAS Collaboration),
\href{doi:10.1038/s41586-024-07824-z}{Nature \textbf{633}, 8030, 542--547 (2024)}

\bibitem{Barr:2024djo}
Alan~J.~Barr, Marco~Fabbrichesi, Roberto~Floreanini et.all,
\href{doi:10.1016/j.ppnp.2024.104134}{Prog. Part. Nucl. Phys. \textbf{139}, 104134 (2024)}

\bibitem{Dong:2023xiw}
Zhongtian~Dong, Dorival~Gon{\c{c}}alves, Kyoungchul~Kong and Alberto~Navarro,
\href{doi:10.1103/PhysRevD.109.115023}{Phys. Rev. D \textbf{109}, 11, 115023 (2024)}

\bibitem{Achasov:2023gey}
M.~Achasov \textit{et al.},
\href{doi:10.1007/s11467-023-1333-z}{Front. Phys. (Beijing) \textbf{19}, 1, 14701 (2024)}

\bibitem{Jackson:1975qi}
John David~Jackson,
\href{doi:10.1103/RevModPhys.48.417}{Rev. Mod. Phys. \textbf{48}, 417--433 (1976)}

\bibitem{Cao:2024tvz}
X.~Cao, Y.~T.~Liang and R.~G.~Ping,
\href{doi:10.1103/PhysRevD.110.014035}{Phys. Rev. D \textbf{110},1, 014035 (2024)}

\bibitem{Tabakin:1985yv}
Frank~Tabakin and R.~A.~Eisenstein,
\href{doi:10.1103/PhysRevC.31.1857}{Phys. Rev. C \textbf{31}, 1857 (1985)}

\bibitem{BESIII:2011nst}
M.~Ablikim \textit{et al.} (BESIII Collaboration),
\href{doi:10.1103/PhysRevD.84.092006}{Phys. Rev. D \textbf{84}, 092006 (2011)}

\bibitem{Karl:1975jp}
Gabriel~Karl, Sydney~Meshkov and Jonathan~L.~Rosner,
\href{doi:10.1103/PhysRevD.13.1203}{Phys. Rev. D \textbf{13}, 1203 (1976)}

\bibitem{ChenPing2020}
Hong~Chen and Rong-Gang~Ping,
\href{doi:10.1103/PhysRevD.102.016021}{Phys. Rev. D \textbf{102}, 016021 (2020)}

\bibitem{Wootters:1997id}
W.~K.~Wootters,
\href{doi:10.1103/PhysRevLett.80.2245}{Phys. Rev. Lett. \textbf{80}, 2245--2248 (1998)}



\bibitem{Ping:2004sh}
R.~G.~Ping, B.~S.~Zou and H.~C.~Chiang,
\href{doi:10.1140/epja/i2004-10069-9}{Eur. Phys. J. A \textbf{23}, 129--133 (2004)}

\bibitem{Ping:2002}
R.~G.~Ping, H.~C.~Chiang and B.~S.~Zou,
\href{doi:10.1103/PhysRevD.66.054020}{Phys. Rev. D \textbf{66}, 054020 (2002)}

\bibitem{3P:1996yt}
E.~S.~Ackleh, Ted~Barnes and E.~S.~Swanson,
\href{doi:10.1103/PhysRevD.54.6811}{Phys. Rev. D \textbf{54}, 6811--6829 (1996)}

\bibitem{mass:2000qj}
Simon~Capstick and W.~Roberts,
\href{doi:10.1016/S0146-6410(00)00109-5}{Prog. Part. Nucl. Phys. \textbf{45}, S241--S331 (2000)}













\end{thebibliography}

\end{document}